\documentclass[journal]{IEEEtran}
\pdfoutput=1

\ifCLASSINFOpdf
\else
\fi

\usepackage{tikz}
\usepackage{graphicx}
\usepackage{citesort}
\usepackage{amsmath, bm}
\usepackage{amsthm}
\usepackage{amssymb, epsfig}
\usepackage{epstopdf}
\usepackage{subfigure,algorithmic,float}
\usepackage{lipsum}
\usepackage{textcomp}
\usepackage{color}
\usepackage{pgfplots}
\usepackage{filecontents}
\usepackage{xcolor}
\usepackage{mathtools}
\usepackage{color}
\usepackage{stfloats}
\pgfplotsset{compat=newest}
\usetikzlibrary{decorations.markings}
\usetikzlibrary{patterns}
\usetikzlibrary{decorations.pathreplacing,angles,quotes}
\theoremstyle{plain}
\newtheorem{lemma}{Lemma}
\newtheorem{prop}{Proposition}

\newtheorem{remark}{Remark}
\theoremstyle{definition}

\providecommand{\keywords}[1]{\textbf{\textit{Index Terms ---}} #1}

\begin{document}

\title{\Huge{Achieving Covert Wireless Communications \\ Using a Full-Duplex Receiver}}

\author{\large{Khurram~Shahzad, \IEEEmembership{Student Member, IEEE,}  Xiangyun~Zhou, \IEEEmembership{Senior Member, IEEE,} Shihao~Yan, \IEEEmembership{Member, IEEE,} Jinsong~Hu, \IEEEmembership{Student Member, IEEE,} Feng~Shu, \IEEEmembership{Member, IEEE,} and Jun~Li,  \IEEEmembership{Senior Member, IEEE}}

\thanks{K. Shahzad and X. Zhou are with the Research School of Engineering, Australian National University, Canberra, ACT, Australia (Emails: \{khurram.shahzad, xiangyun.zhou\}@anu.edu.au).}
\thanks{S. Yan is with the School of Engineering, Faculty of Science and Engineering, Macquarie University, Sydney, NSW, Australia (Email: shihao.yan@mq.edu.au).}
\thanks{J. Hu, F. Shu, and J. Li are with the School of Electronic and Optical Engineering, Nanjing University of Science and Technology, Nanjing, China (Emails: \{jinsong\textunderscore hu, shufeng, jun.li\}@njust.edu.cn).}
\thanks{Part of this work was presented at the IEEE International Conference on Communications (ICC'18) \cite{fd_icc_2018}.}
}

\markboth{IEEE Transactions on Wireless Communications}{Shahzad \MakeLowercase{\textit{et al.}}: Achieving Covert Wireless Communications Using a Full-Duplex Receiver}
\maketitle


\begin{abstract}
Covert communications hide the transmission of a message from a watchful adversary while ensuring a certain decoding performance at the receiver. In this work, a wireless communication system under fading channels is considered where covertness is achieved by using a full-duplex (FD) receiver. More precisely, the receiver of covert information generates artificial noise with a varying power causing uncertainty at the adversary, Willie, regarding the statistics of the received signals. Given that Willie's optimal detector is a threshold test on the received power, we derive a closed-form expression for the optimal detection performance of Willie averaged over the fading channel realizations. Furthermore, we provide guidelines for the optimal choice of artificial noise power range, and the optimal transmission probability of covert information to maximize the detection errors at Willie. Our analysis shows that the transmission of artificial noise, although causes self-interference, provides the opportunity of achieving covertness but its transmit power levels need to be managed carefully. We also demonstrate that the prior transmission probability of $0.5$ is not always the best choice for achieving the maximum possible covertness, when the covert transmission probability and artificial noise power can be jointly optimized.
\end{abstract}



\IEEEpeerreviewmaketitle

\section{Introduction}
\subsection{Background}
The wireless air interface is open and accessible to both legitimate and illegitimate users. This creates reasonable concerns over the security and privacy of information transmitted over the air. The recent remarkable increase in the amount of information conveyed using the wireless medium has spurred an interest in both research and academic communities regarding the development of new mechanisms, enhancing the privacy and integrity of wirelessly transmitted data. In recent years, physical layer security \cite{bloch_book,sean_book} has emerged as an alternative to traditional cryptographic ways of securing wireless information, where the mechanisms of key exchange and distribution impose varied challenges, especially in dynamic network environments. Physical layer security techniques exploit the uncertainties and lack of predictability of the wireless channel, minimizing the information obtained by an unauthorized eavesdropper. Under the varied circumstances where users communicate over the wireless medium, situations exist where not only the privacy and integrity of the information are important, but the users may also wish to avert any invigilation, hiding the very existence of their communication. Such situations, although, commonplace in military applications, are now also arising in non-military applications, relating to civil unrest and even monitoring of people's daily activities. Thus, catering for such security concerns is a need of the moment and motivates the recent interest in covert communications \cite{commag15bash,bloch_TIT}.

Covert communications intend to obscure the existence of any wireless transmission from a watchful adversary, referred to as \textit{Willie} in recent literature of covert communications, while guaranteeing a certain decoding performance at the intended receiver. The low probability of detection (LPD) communications have drawn significant research attention and are materializing as a promising prospect for shielding the future wireless communication networks from unapproved probing and access. In this regard, the fundamental limits of covert communications were established in \cite{bash_jsac}, presenting a square root limit on the amount of information transmitted reliably and with low probability of detection over additive white Gaussian noise (AWGN) channels. This work has been further extended to binary symmetric channels (BSCs) \cite{jaggi_ISIT_13}, discrete memoryless channels (DMCs) \cite{wang_TIT} and multiple access channels (MACs) \cite{bloch_ISIT_16}. Although under the square root law, the average number of covert bits per channel use reaches asymptotically zero, a positive covert rate has been shown to be achievable in a number of cases. These include the situations of Willie's uncertainty in the knowledge of noise power \cite{lee_undetect,goeckel_noise_2016,biao_cc,hu2018covert}, Willie's ignorance of transmission time \cite{ignorance_2016}, and presence of a continuously transmitting jammer in the environment \cite{tamara_jammer_2017}. The case when additional \textit{friendly} nodes generating artificial noise are present in the environment, causing confusion at Willie regarding the received signal statistics, is presented in \cite{soltani2017covert}, while analysis of covert transmissions under finite blocklengths, imperfect channel state information and in one-way relay networks is presented in \cite{shihao_finite_2017,yan2018delay}, \cite{shahzad_vtc} and \cite{hu2018covert_relay}, respectively.

\subsection{Our Approach and Contribution}
In this work, we make use of a full-duplex (FD) receiver to achieve covert communication. Specifically, the FD receiver generates artificial noise (AN) with a randomized transmit power, causing a deliberate confusion and affecting the decisions at Willie regarding the presence of any covert transmissions. Although the use of AN and jamming signals for enhancing physical layer security has been widely advocated in the literature
\cite[and references therein] {yan2017three,phy_fd_jam_2013,cumanan2017physical,ghogho_an_2012,yan2016artificial}, to the best of our knowledge, it has not been studied before in the context of covert communications. The use of a FD receiver generating AN provides a cover for the covert transmission, and offers a multitude of benefits as compared to the use of a separate, independent jammer. Being equipped with an FD receiver, we can exercise a better control over the power used for transmitting AN, hence a better management of system resources to achieve the said purpose of security is achievable. Furthermore, while Willie will face a strong interference, the self-interference at the FD receiver can be greatly suppressed by the well-developed self-interference cancellation techniques \cite{passive_fd_2014,zhu2016physical}, providing a significant advantage to the covert communication pair.

In our considered scenario, covert transmissions can occur in multiple blocks of time, and Willie is performing the detection on a block-to-block basis. In this case, the \textit{a priori} transmission probability becomes an interesting and important parameter affecting Willie's detection performance as well as the overall throughput of covert communications. A general assumption in the literature regarding the \textit{a priori} probability of covert transmission is that there is a $50\%$ chance that transmission occurs in a block of interest. This assumption is understood as a good choice for covertness, since it renders Willie's knowledge of Alice's transmission uninformative and is equivalent to assuming that Willie has no prior knowledge on whether Alice transmits or not \cite{bash_jsac,ignorance_2016}. We show in this work that an \textit{a priori} probability of $0.5$ is not always the best choice in our considered scenario; rather a joint adjustment of this probability with other system parameters can offer a better covert performance.

The main contributions of this work are summarized as follows:

\begin{itemize}
\item We show that the use of an FD receiver is an effective way of achieving covert communication over fading wireless channels, where the FD receiver is designed to transmit AN with varying power to cause confusion at Willie.

\item Under the assumption of a radiometer (power-detector) at Willie, we analytically derive the optimal detection threshold of Willie's radiometer and obtain its optimal detection performance in terms of the minimum detection error probability.

\item For a given covert rate requirement, we provide the design guidelines on optimal choices for the range of AN transmit power at the FD receiver and the optimal \textit{a priori} probability of covert transmission in order to maximize the expected detection error probability at Willie.

\item Our analysis reveals that an \textit{a priori} transmission probability of $0.5$ is not always the best choice. Increasing this transmission probability beyond $0.5$ gives more room to increase the AN transmit power for maintaining the same rate requirement. Thus whether to allow such a change in the transmission probability can be the difference between achieving strong covertness and achieving almost no covertness at all.
\end{itemize}

\subsection{Related Works}
Our work is closely related to \cite{tamara_jammer_2017}, where a jammer is assumed to be present in the environment. Although the jammer does not closely coordinate with the covert transmitter, it is allowed to transmit continuously and the received power at Willie due to the jammer changes randomly from slot to slot. In this case, the covert communication pair has no control over the jammer's transmit power level. In contrast, although we also consider randomizing the AN power in each slot, our focus is on optimizing the AN transmit power range, since this choice affects the information decoding at the intended receiver through self-interference. This important optimization is made possible because the AN is transmitted by the FD receiver, and hence, controllable by the covert communication pair. Moreover, instead of satisfying a given covertness constraint, we present our analysis on the choice of AN transmit power range to achieve the maximum possible covertness while meeting a given rate requirement.

Furthermore, performance of communication systems with randomly distributed interferers has been studied extensively in the literature \cite{pinto2010communication1,pinto2010communication2,shabsigh2016quantifying}. More recently, a study on covert communications in the presence of a Poisson distributed field of interferers has been presented in \cite{he2017covert}, where leveraging the total received interference, the effect of density and transmit powers of the interferers on the covert throughput is analyzed. Our work differs from \cite{he2017covert} in that we consider AN generated by the FD receiver, hence allowing design and optimization of AN power with other design parameters. Thus, while the authors in \cite{he2017covert} study the covert performance for a given interference scenario, we take a design approach and provide guidelines on the optimal choice of parameters for achieving covertness.

\subsection{Organization}
The rest of this paper is organized as follows: Section II details our communication scenario, proposed scheme and the assumptions used in this paper. Section III explains Willie's approach for detection of any covert transmissions, deriving the conditions for possibility of any covert communications and the optimal settings at Willie. Using the knowledge of Willie's approach, Section IV discusses the parameters that affect the achievable performance of the proposed communication scheme while Section V addresses the optimal design of all the system parameters that we have in our control to achieve the best possible performance in covertness. Section VI provides numerical results validating our analysis and provides further insights on the impact of AN and priori probabilities, and finally, Section VII draws some concluding remarks.

\section{System Model}
\subsection{Communication Scenario}
\begin{figure}[t]
\centering
	\includegraphics[width=\linewidth]{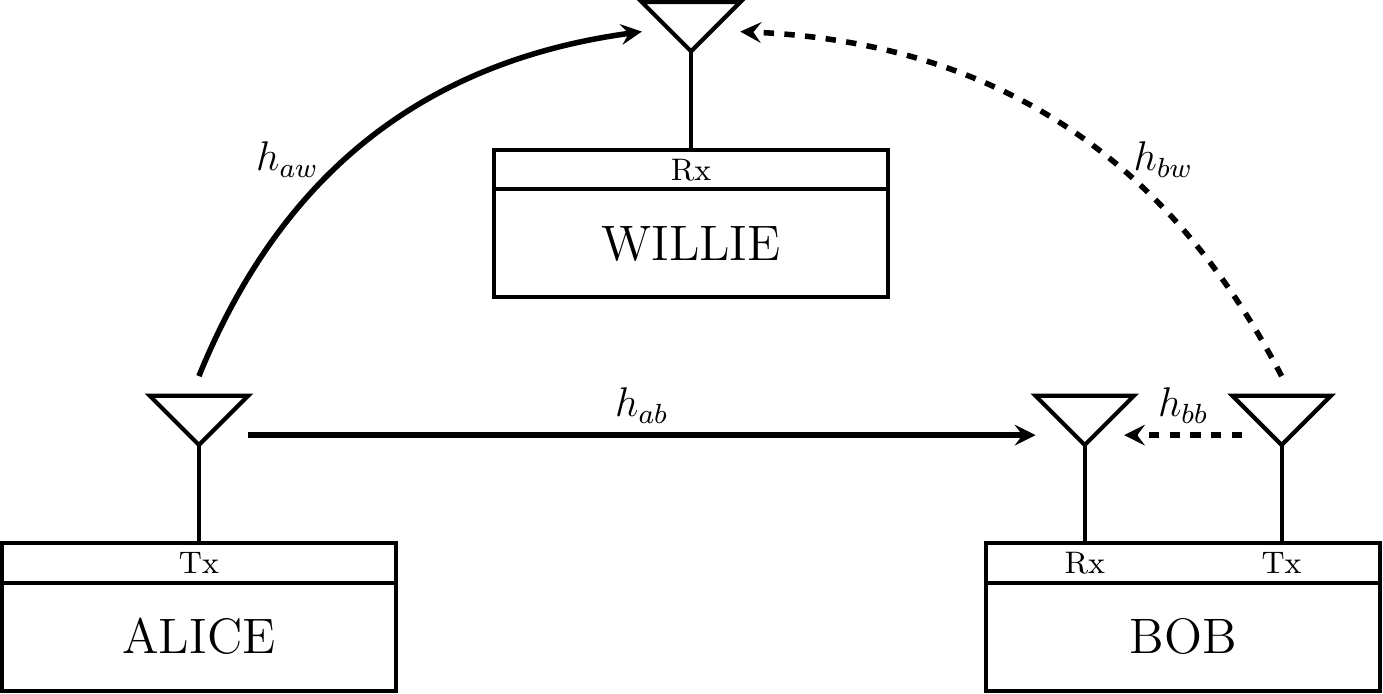}
	\caption{Covert communications model under consideration with a FD receiver.}
	\label{system_model_fd}
\end{figure}

A covert wireless communication system is considered, as shown in Fig. \ref{system_model_fd}, where a transmitter (Alice) possesses sensitive information that needs to be sent to an information receiver (Bob). Bob operates in full-duplex mode, and Alice seeks to transmit covertly to Bob with the aid of artificial noise (AN) generated by Bob. Under these circumstances, an adversary (Willie) silently listens to the communication environment and tries to detect any covert transmission from Alice to Bob. We use the subscripts $a$, $b$ and $w$ to represent the terms associated with Alice, Bob and Willie, respectively. It is assumed that Willie has complete knowledge of the carrier frequency of any transmissions, associated antenna gains and the distances between all the nodes.

A communication slot is defined as a block of time over which the transmission of a message from Alice to Bob is completed. Each slot contains $n$ symbol periods and we assume that $n$ is large enough, i.e., $n\rightarrow\infty$. The slot boundaries are perfectly synchronized among all the users, and we consider fading wireless channels where the channel coefficients remain constant in one slot, changing independently from one slot to another, i.e., quasi-static Rayleigh fading channels are considered. The channel between any two users $i$ and $j$ is represented by $h_{ij}$, where the channel gain is assumed to encompass the combined antenna gain of transmit/receive antennae and the distance between the two users as well. The mean of $|h_{ij}|^2$ over different communication slots is denoted by $1/\lambda_{ij}$, where subscript $ij$ can be $ab$, $aw$, $bw$ or $bb$. Hence, the Alice-Bob, Alice-Willie and Bob-Willie channels are denoted by $h_{ab}$, $h_{aw}$, and $h_{bw}$, respectively, while the self-interference channel of Bob is denoted by $h_{bb}$. We note here that $h_{bb}$ is the loop interference channel at Bob and is modelled via the Rayleigh fading distribution under the assumption that any line-of-sight component is efficiently reduced by antenna isolation and the major effect comes from scattering \cite{ngo2014multipair}. Regarding the channel knowledge, it is assumed that Bob knows $h_{ab}$, while Willie possesses complete knowledge of $h_{aw}$ and $h_{bw}$. Here, the availability of knowledge regarding $h_{aw}$ and $h_{bw}$ at Willie represents the worst case scenario from the perspective of covert communication design.

The complex additive Gaussian noise at Bob and Willie's receiver is denoted by $n_b \sim \mathcal{CN}(0,\sigma_b^2)$ and $n_w \sim \mathcal{CN}(0,\sigma_w^2)$, respectively. Each of Alice and Willie is equipped with a single antenna, while apart from a receiving antenna, Bob also has an additional antenna for the transmission of AN.  Due to its full-duplex nature, Bob suffers from self-interference, causing a degradation in the signal-to-noise ratio (SNR) of the message signal received from Alice \cite{sabharwal2014band,duarte_fd_2012}. Since the generated AN signal is known to Bob, the self-interfering signal, acting as noise for Bob's receiver, can be rebuilt and eliminated up to a certain extent by using efficient techniques of self-interference cancellation \cite{passive_fd_2014,katti_fd_2013}. However, owing to computational limitations and practical non-idealities, we assume that perfect cancellation of self-interference is not achievable \cite{krikidis2012full}. The self-interference cancellation coefficient is denoted by $\phi$, where $0 < \phi \leq 1$ corresponds to different cancellation levels of interfering AN signals. The residual interfering link is also modelled as Rayleigh fading channel, following a common assumption in the literature \cite{ngo2014multipair,riihonen2011mitigation}.

The transmit power of Alice and Bob is denoted by $P_a$ and $P_b$, respectively. When Alice transmits, the signal received at Bob for each symbol period is given by
\begin{equation}
y_b(i) = \sqrt{P_a}h_{ab}x_a(i) + \sqrt{\phi P_b}h_{bb}x_b(i) + n_b(i) ,
\end{equation}
where $i=1,\dots,n$ represents the symbol index. Here, $x_a$ and $x_b$ represent the signals transmitted by Alice and Bob, respectively, satisfying $\mathbb{E}[x_a(i) x_a^{\dagger}(i) ] = 1$ and  $\mathbb{E}[x_b(i) x_b^{\dagger}(i)] = 1$. We also consider an average power constraint on Bob's transmit power, denoted by $P_{\text{avg}}$. We follow the common assumption that a secret of sufficient length is shared between Alice and Bob \cite{bash_jsac,tamara_jammer_2017}, which while unknown to Willie, enables Bob to know Alice's strategy. Employing random coding arguments, Alice generates codewords of length $n$, by independently drawing symbols from a zero-mean complex Gaussian distribution with unit variance. Here, each codebook is known to Alice and Bob and is used only once. When Alice transmits in a slot, she selects the codeword corresponding to her message and transmits the resulting sequence.

\subsection{Proposed Transmission Scheme}
We propose a communication scheme that allows Bob to receive Alice's transmission covertly, exploiting an AN signal generated by Bob, where the transmit power of this AN changes from slot to slot. Alice's transmit power, $P_a$, is fixed and publicly known by both Willie and Bob. On the other hand, $P_b$, defined as the average power used by Bob for AN transmission in a given slot, changes from slot to slot, following a continuous uniform distribution over the interval $\left[P_{\text{min}}, P_{\text{max}} \right]$, having a probability density function (pdf) given by
\begin{equation}
\begin{aligned}
f_{P_b} (p) =
\begin{cases}
\frac{1}{P_{\text{max}}-P_{\text{min}}}, &\text{if} \quad P_{\text{min}} \leq p \leq P_{\text{max}} \\
0, &\text{otherwise}.
\end{cases}
\end{aligned}
\end{equation}

\begin{figure}[t]
\centering
	\includegraphics[width=\linewidth]{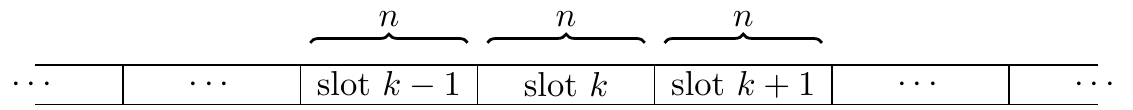}
	\caption{Willie's observation model, presented as a slotted channel use by Alice. Each slot contains $n$ symbol periods and there is a certain \textit{a priori} probability, $\pi_1$, of Alice's communication to Bob in each slot.}
	\label{w_obs_model}
\end{figure}

It should be noted here that in the proposed scheme, Bob transmits the AN signal continuously, regardless of whether or not Alice transmits in a given slot. In this work, we address the covertness regarding Alice's existence and message transmission to Bob, whereas we are not trying to hide the existence of Bob. Apart from being the information receiver, Bob also plays the role of a cooperative jammer and hence his presence is known to Willie. Willie has complete knowledge of Bob's AN power distribution, but the exact power used by Bob in a given slot is unknown to Willie. Due to the independent Gaussian nature of Alice and Bob's transmitted signals and Willie's receiver noise, the signal received at Willie is always Gaussian in any slot, regardless of whether Alice transmits or not, thus Willie can not make use of any difference in the distribution type of the received signal for detection purposes. Since Willie knows the channels $h_{aw}$ and $h_{bw}$ in any slot under consideration, for a constant transmit power from Bob, it is straightforward for him to raise an alarm when an additional power from Alice is received. By introducing randomness in Bob's transmit power, we create an uncertainty at Willie, causing confusion as to whether the increase in received signal power is due to Alice's covert transmission or merely a change in Bob's AN power. This effectively creates an artificial fading for Willie \cite{an_af_2014}, despite the fact that he has the perfect channel knowledge for both Alice and Bob.

\subsection{Detection at Willie}
As described earlier, it is assumed that Willie is unaware of the exact transmit power used by Bob in each slot, although the distribution of Bob's AN transmit power is known to Willie. Also, Willie has full knowledge of the associated antenna gains, distances among all the nodes and his own receiver's noise variance. We consider a covert communications scenario that spans a large number of slots, and there is a possibility of transmission by Alice in each slot. Due to the shared secret between Alice and Bob, this constitutes a form of a time-hopping system. Here, Willie looks to make a decision regarding Alice's transmission in each slot as he is interested in knowing for each individual slot that whether Alice transmitted or not. This means that Willie is not only interested in ``\textit{whether}" a transmission happens but also ``\textit{when}" it happens, i.e., in which slot. Note that if Willie is only interested in whether transmission happens but not in when it happens, he needs to make only a single decision after observing all slots. Such a scenario has been considered in \cite{ignorance_2016}, where the slot selection is kept secret from Willie and he looks to make a single decision regarding Alice's transmission over all possible slots.

The knowledge of ``\textit{when}" a transmission happens not only improves upon Willie's effectiveness in detecting covert transmission, but also gives him the capability of taking an action at the required time rather than waiting for the end of observation interval before intervening (although the corresponding action by Willie is beyond the scope of this work). Consider a scenario where Alice and Bob agree upon a certain ``pattern" in choosing the slots over which covert messages are sent. Once Willie is able to detect  the pattern based on his per slot decisions, it becomes easier for him to efficiently predict the slots over which future transmissions will happen\footnote{Although the proposed scheme will help Willie in being able to predict any such pattern, this prediction is beyond the scope of this work and is thus not considered here.}.

\subsection{Priors and Performance Metrics}
Willie faces a decision as to whether or not Alice sent any covert information to Bob. As a result, Willie faces a binary hypothesis testing problem. The null hypothesis, $H_0$, states that Alice did not transmit while the alternative hypothesis, $H_1$, states that Alice did transmit, sending covert information to Bob. We define the probability of false alarm (or Type I error) as the probability that Willie makes a decision in favor of $H_1$, while $H_0$ is true, denoted by $\mathbb{P}_{FA}$. Similarly, the probability of missed detection (or Type II error) is defined as the probability of Willie making a decision in favor of $H_0$, while $H_1$ is true, and is denoted by $\mathbb{P}_{MD}$. We denote by $\pi_0$ and $\pi_1$ the \textit{a priori} probabilities of hypothesis $H_0$ and $H_1$, respectively. The detection error probability at Willie is given by
\begin{equation}\label{eq11}
\mathbb{P}_E=\pi_0\mathbb{P}_{FA} + \pi_1\mathbb{P}_{MD} ,
\end{equation}
which serves as a measure of covertness. In the recent literature, the assumption of both hypotheses being presented with an equal \textit{a priori} probability has been widely adopted \cite{lee_undetect,ignorance_2016}. The knowledge of \textit{a priori} probabilities helps Willie improve his detection performance \cite[Fact $4$] {bash_jsac}, as his assumption of $\pi_0 = \pi_1 = \frac{1}{2}$ implies that his observations are of little use to him and his decisions are akin to a random guess about the transmission state of Alice. In this work, we instead consider general, i.e., not necessarily equal priors, and assume that Willie happens to know them. Since $\mathbb{P}_E \leq \min \left(\pi_0, \pi_1 \right)$, achieving covert communication guarantees that $\mathbb{P}_E$ is in close proximity of $\min \left(\pi_0, \pi_1 \right)$.

\section{Detection Scheme at Willie}
The signals received at Willie under the two possible hypotheses for each symbol period are given by
\begin{equation}
\begin{aligned}
y_w(i) =
\begin{cases}
\sqrt{P_a}h_{aw}x_a(i) + \sqrt{P_b}h_{bw}x_b(i) + n_w(i),  &\text{Alice transmits} \\
 \sqrt{P_b}h_{bw}x_b(i) + n_w(i), &\text{else}.
\end{cases}
\end{aligned}
\end{equation}
From the independent and identically distributed (i.i.d.) nature of Willie's received vector, $\bm{y_w}$, each element of $\bm{y_w}$, i.e., $y_w(i)$ has a distribution given by
\begin{equation}
\begin{aligned}
\begin{cases}
\mathcal{CN} (0, |h_{aw}|^2P_a + |h_{bw}|^2P_b + \sigma_w^2),   & \text{Alice transmits}   \\
\mathcal{CN} (0, |h_{bw}|^2P_b + \sigma_w^2),& \text{else}.
\end{cases}
\end{aligned}
\end{equation}
We note that while the distribution of $P_b$ is known to Willie, its value in a given slot is not known. Based on his observation vector $\bm{y_w} = [y_w(1), \dots, y_w(n)]$, Willie has to make a decision regarding Alice's actions in each communication slot. We assume that Willie uses a radiometer as his detector \cite{tamara_jammer_2017,lee_undetect} due to its low complexity and ease of implementation. When Willie has the statistical knowledge of his observations, this assumption is justified and the optimality of radiometer can be proved along the same lines as the proof of Lemma $3$ in \cite{tamara_jammer_2017} using Fisher-Neyman Factorization Theorem \cite{degroot2012probability} and Likelihood Ratio Ordering concepts \cite{stoch_order}. While adopting a radiometer, the total received power at Willie, $\sum_{i=1}^{n}|y_w(i)|^2$ is a sufficient statistic for Willie's test. Since any one-to-one transformation of a sufficient statistic is also sufficient, the term $1/n \sum_{i=1}^{n} |y_w(i)|^2$ is also a sufficient statistic. Thus Willie conducts a threshold test on the average power received in a slot, given by

\begin{equation}
P_w \underset{D_0}{\overset{D_1}{\gtrless}} \gamma ,
\end{equation}
where $P_w = 1/n \sum_{i=1}^{n} |y_w(i)|^2$ is the average power received at Willie in a slot, $D_0$ and $D_1$ are defined as the events that Willie makes a decision in the favor of $H_0$ and $H_1$, respectively, and $\gamma$ is Willie's detector threshold, which can be optimized to minimize the detection error probability. The average power at Willie in a slot under hypothesis $H_0$ is given by
 \begin{equation}
 \begin{aligned}
 P_w (H_0) &=  \underset{n \rightarrow \infty}{\lim} \left(|h_{bw}|^2 P_b + \sigma_w^2 \right) \chi_{2n}^{2} / n \\
 &= |h_{bw}|^2 P_b + \sigma_w^2,
 \end{aligned}
 \end{equation}
where $\chi_{2n}^{2}$ represents a chi-squared random variable with $2n$ degrees of freedom and from the Strong Law of Large Numbers, we know that $\chi_{2n}^{2 } / n \rightarrow 1$ almost surely. Similarly, the average power at Willie in a slot under hypothesis $H_1$ is
\begin{equation}
P_w (H_1) =  |h_{bw}|^2 P_b + |h_{aw}|^2 P_a + \sigma_w^2.
  \end{equation}

We first analyze the condition under which Willie has non-zero probability of making detection errors and based on that, we find the optimal setting for Willie's detector threshold. It should be noted here that the analysis of Willie's detection error probability presented in the following proposition is for given channel realizations as Willie possesses the full knowledge of his channel from Alice and Bob.

\begin{prop}
Willie has a non-zero probability of making detection errors when:
\begin{equation}\label{prop1.1}
\frac{|h_{aw}|^2}{|h_{bw}|^2} \leq \frac{P_{\text{max}} - P_{\text{min}}}{P_a} .
\end{equation}
When (\ref{prop1.1}) holds, the optimal choice for Willie's detector's threshold is
\begin{equation}\label{3.9}
\begin{aligned}
\gamma^* =
\begin{cases}
|h_{bw}|^2 P_{\text{min}} + |h_{aw}|^2 P_{\text{a}} + \sigma_w^2, &\text{if} \quad \pi_1 \geq \pi_0 \\
|h_{bw}|^2 P_{\text{max}} + \sigma_w^2, &\text{otherwise},
\end{cases}
\end{aligned}
\end{equation}
and the corresponding minimum detection error probability at Willie is given by
\begin{equation}\label{3.10}
\begin{aligned}
\mathbb{P}_{E}^* =
\begin{cases}
\pi_0 \left[1 - \frac{|h_{aw}|^2 P_a}{|h_{bw}|^2 \left(P_{\text{max}} - P_{\text{min}}  \right)}  \right], & \text{if} \quad \pi_1 \geq \pi_0 \\
\pi_1 \left[1 - \frac{|h_{aw}|^2 P_a}{|h_{bw}|^2 \left(P_{\text{max}} - P_{\text{min}}  \right)}  \right], & \text{otherwise} .
\end{cases}
\end{aligned}
\end{equation}
\end{prop}

\begin{IEEEproof}
The detailed proof is provided in Appendix A.
\end{IEEEproof}

\begin{remark}
From Proposition $1$, when (\ref{prop1.1}) does not hold, Willie will have zero probability of making a detection error by setting the threshold $\gamma$ in the interval $|h_{bw}|^2 P_{\text{max}}+\sigma_w^2 < \gamma \leq |h_{bw}|^2 P_{\text{min}} + |h_{aw}|^2 P_\text{a}+\sigma_w^2$. We also note here that although Willie's receiver noise variance, $\sigma_w^2$, is required for the calculation of the optimal threshold for Willie's detector, its value does not affect the minimum detection error probability at Willie. This can be attributed to the fact that as $n \rightarrow \infty$, there is no uncertainty at Willie regarding the noise statistics and hence it does not contribute to an increase or decrease in the detection error probability at Willie.
\end{remark}

\section{Performance of Covert Communication}
In this section, we present those system metrics which affect the performance of our proposed covert transmission scheme. We note that the square root law presented by Bash et al. \cite{bash_jsac} holds given Willie has perfect statistical knowledge of the test statistics. It has been shown in prior works \cite{lee_undetect,tamara_jammer_2017,biao_cc} that uncertainties present (or intentionally introduced) in the test statistics under both the null and alternative hypotheses at Willie result in a positive rate. In this work, the randomness in Bob's transmit power introduces the required uncertainty at Willie, and hence we are able to achieve a positive covert rate. Here, we first calculate the outage probability for the transmission from Alice to Bob, and then present a measure that helps in quantifying the performance of our presented covert scheme.

\subsection{Transmission Outage Probability from Alice to Bob}
The signal-to-interference-plus-noise ratio (SINR) at Bob, in case Alice transmits,  is given by
\begin{equation}\label{4.1}
\text{SINR}_b = \frac{|h_{ab}|^2 P_a}{\phi |h_{bb}|^2 P_b + \sigma_b^2} .
\end{equation}
We assume a pre-determined rate from Alice to Bob, and denote it by $R_{ab}$. Due to the random nature of $h_{ab}$, $h_{bb}$ and $P_b$, a transmission outage from Alice to Bob occurs when $C_{ab} < R_{ab}$, where $C_{ab}$ is the channel capacity from Alice to Bob.

\begin{lemma}
The transmission outage probability from Alice to Bob is given by
\begin{equation}\label{4.2}
\delta_{ab} = 1 - \frac{\lambda_{bb}\exp(- \lambda_{ab}\mu\sigma_b^2)}{(P_{\text{max}}-P_{\text{min}}) \lambda_{ab}\phi\mu } \ln \left[ \frac{\lambda_{bb}+\lambda_{ab}\phi\mu P_{\text{max}}}{\lambda_{bb}+\lambda_{ab}\phi\mu P_{\text{min}}}     \right] ,
\end{equation}
\end{lemma}
where $\mu \triangleq \left(2^{R_{ab}} -1 \right) / P_a$.
\begin{IEEEproof}
From the definition of transmission outage probability, we have
\begin{equation}\label{4.3}
\begin{aligned}
\delta_{ab} &= \mathcal{P}\left[ C_{ab} < R_{ab} \right] \\
&= \mathcal{P} \left[  \frac{|h_{ab}|^2 P_a}{\phi |h_{bb}|^2 P_b + \sigma_b^2} < 2^{R_{ab}} - 1 \right] \\
&= \int_{P_{\text{min}}}^{P_{\text{max}}} \int_{0}^{\infty} \int_{0}^{\mu (\phi |h_{bb}|^2 P_{b}+\sigma_b^2)} f_{|h_{ab}|^2}(x) f_{|h_{bb}|^2}(y) f_{P_b}(z) \\ & \hspace{6cm}   \times \mathrm{d}x \: \mathrm{d}y \: \mathrm{d}z \\
&= \int_{P_{\text{min}}}^{P_{\text{max}}} \int_{0}^{\infty}  \Big[ 1 - \exp \left(-\lambda_{ab}\mu (\phi |h_{bb}|^2 P_b + \sigma_b^2)\right)   \Big]  \\ & \hspace{4.7cm} \times  f_{|h_{bb}|^2}(y) f_{P_b}(z) \: \mathrm{d}y \: \mathrm{d}z \\
&= \int_{P_{\text{min}}}^{P_{\text{max}}} \left[1 - \frac{\lambda_{bb}\exp \left(-\lambda_{ab}\mu \sigma_b^2 \right)}{\lambda_{bb}+\lambda_{ab}\mu \phi P_b} \right] f_{P_b}(z) \mathrm{d}z \\
&= 1 - \frac{1}{P_{\text{max}} - P_{\text{min}}} \int_{P_{\text{min}}}^{P_{\text{max}}} \left[ \frac{\lambda_{bb}\exp \left(-\lambda_{ab}\mu \sigma_b^2 \right)}{\lambda_{bb}+\lambda_{ab}\mu \phi z} \right] \mathrm{d}z ,
\end{aligned}
\end{equation}
and using the solution from \cite{gradshteyn2014table} for the general form of integral $ \int \frac{A}{B+Cx} \mathrm{d}x = \frac{A \log (B+Cx)}{C}$ for the second term gives the desired result.
\end{IEEEproof}

\subsection{Expected Detection Error Probability at Willie}
Since Alice and Bob are unaware of their instantaneous channel to Willie, we consider the expected value of detection error probability at Willie, $\mathbb{P}_E^*$, over all possible realizations of $h_{aw}$ and $h_{bw}$ as the measure of covertness from the viewpoint of Alice and Bob, and this expected detection error probability at Willie is denoted by $\overline{\mathbb{P}_{E}^*}$.

\begin{lemma}
Under the optimal detection threshold setting, the expected detection error probability at Willie is given by
\begin{equation}\label{4.8}
\begin{aligned}
\overline{\mathbb{P}_{E}^*} =
\begin{cases}
\pi_0 \left[ 1 + t \ln t - t^2    \right], & \text{if} \quad \pi_1 \geq \pi_0 \\
\pi_1 \left[ 1 + t \ln t - t^2    \right], & \text{otherwise} ,
\end{cases}
\end{aligned}
\end{equation}
where $t \triangleq ( \lambda_{bw}P_a) / \left[ \lambda_{bw}P_a + \lambda_{aw}\left( P_{\text{max}}-P_{\text{min}}  \right)\right] $.
\end{lemma}

\begin{IEEEproof}
For the case of $\pi_1 \geq \pi_0$, and under the condition of Willie making detection errors, given by $|h_{bw}|^2 P_{\text{max}}+\sigma_w^2 \geq |h_{bw}|^2 P_{\text{min}} + |h_{aw}|^2 P_a+\sigma_w^2$, we have
\begin{equation}\label{4.9}
\begin{aligned}
\overline{\mathbb{P}_{E}^*} = \pi_0 \bigg\{ \int_{0}^{\infty} \int_{0}^{\frac{|h_{bw}|^2 \left( P_{\text{max}} - P_{\text{min}} \right)}{P_a}}  \hspace{-1cm} \mathbb{P}_{E}^*  \quad  f_{|h_{aw}|^2}(x) f_{|h_{bw}|^2}(y) \mathrm{d}x \mathrm{d}y     \bigg \} ,
\end{aligned}
\end{equation}
which, using the law of total expectation, can also be written as
\begin{equation}\label{4.9}
\begin{aligned}
\overline{\mathbb{P}_{E}^*} &= \pi_0 \bigg\{  \mathcal{P}\left[|h_{aw}|^2 \leq \frac{|h_{bw}|^2 \left( P_{\text{max}} - P_{\text{min}} \right)}{P_a} \right] \\ & \hspace{1cm} \times \mathbb{E} \left[ \mathbb{P}_{E}^* \Big| |h_{aw}|^2 \leq \frac{|h_{bw}|^2 \left( P_{\text{max}} - P_{\text{min}} \right)}{P_a} \right]  \bigg \} ,
\end{aligned}
\end{equation}
where
\begin{equation}\label{4.6}
\begin{aligned}
\mathcal{P} &\left[|h_{aw}|^2 \leq \frac{|h_{bw}|^2 \left( P_{\text{max}} - P_{\text{min}} \right)}{P_a} \right]  \\
&= \int_{0}^{\infty} \int_{0}^{\frac{|h_{bw}|^2 \left( P_{\text{max}} - P_{\text{min}} \right)}{P_a}} f_{|h_{aw}|^2}(x) f_{|h_{bw}|^2}(y) \mathrm{d}x \mathrm{d}y \\
&= \int_{0}^{\infty} \left[ 1 -  \exp \left(- \frac{\lambda_{aw}\left( P_{\text{max}} - P_{\text{min}} \right)y}{P_a}  \right)  \right] \times \\ &\qquad \qquad \qquad \qquad \qquad \qquad \lambda_{bw} \exp \left(- \lambda_{bw} y   \right) \mathrm{d}y \\
&=  \frac{\lambda_{aw}\left( P_{\text{max}} - P_{\text{min}} \right)}{\lambda_{bw}P_a + \lambda_{aw}\left( P_{\text{max}} - P_{\text{min}} \right)} ,
\end{aligned}
\end{equation}
and
\begin{equation}\label{4.7}
\begin{aligned}
\mathbb{E} & \left[ \mathbb{P}_{E}^* \Big| [|h_{aw}|^2 \leq \frac{|h_{bw}|^2 \left( P_{\text{max}} - P_{\text{min}} \right)}{P_a} \right] \\
&= 1 - \frac{P_a}{\left( P_{\text{max}} - P_{\text{min}} \right)}  \\
& \hspace{1.5cm} \times \mathbb{E} \left[ \frac{|h_{aw}|^2}{|h_{bw}|^2} \: \Big| \:  |h_{aw}|^2 \leq \frac{|h_{bw}|^2 \left( P_{\text{max}} - P_{\text{min}} \right)}{P_a}  \right] \\
&= 1 - \frac{P_a}{\left( P_{\text{max}} - P_{\text{min}} \right)} \int_{0}^{\infty} \int_{0}^{\frac{|h_{bw}|^2 \left( P_{\text{max}} - P_{\text{min}} \right)}{P_a}} \frac{x}{y} \\
& \hspace{4cm} \times f_{|h_{aw}|^2}(x) f_{|h_{bw}|^2}(y) \mathrm{d}x \mathrm{d}y \\
&= 1 - \frac{\lambda_{bw}P_a}{\lambda_{aw}\left( P_{\text{max}} - P_{\text{min}} \right)} \bigg[ \ln \left(1 + \frac{\lambda_{aw}\left( P_{\text{max}} - P_{\text{min}} \right)}{\lambda_{bw}P_a} \right) \\
& \hspace{3cm}- \frac{\lambda_{aw}\left( P_{\text{max}} - P_{\text{min}} \right)}{\lambda_{bw}P_a + \lambda_{aw}\left( P_{\text{max}} - P_{\text{min}} \right)}  \bigg] ,
\end{aligned}
\end{equation}
and putting in these expressions into (\ref{4.9}) gives the desired result.

The case for $\pi_0 > \pi_1$ follows along the same lines, hence concluding the proof.
\end{IEEEproof}

\begin{remark}
We make a few observations regarding the effect of $P_{\text{max}}$ and $P_a$ on Willie's detection performance. Firstly, as $P_{\text{max}} \rightarrow \infty$, the probability of Willie making detection errors approaches $\pi_0$ or $\pi_1$, in respective cases, which represents the maximum of $\overline{\mathbb{P}_{E}^*}$. Secondly, if Alice's transmit power $P_a \rightarrow \infty$, then $t \rightarrow 1$ and $\overline{\mathbb{P}_{E}^*} \rightarrow 0$. Thus for a given set of $\{P_{\text{min}}, P_{\text{max}} \}$, Alice can be ``loud" enough to be heard by Willie.
\end{remark}

\section{Covert Communication Design}
In majority of the recent literature in covert communications, the detection error probability is used to measure the level of covertness under the assumption of equal priors. However, in this work, we propose a different framework and  instead of putting a constraint on the error probability to achieve a said covertness, we look to maximize it under the given system model. Hence, from Alice and Bob's perspective, the objective is to achieve the best possible covertness in transmission, while being subject to an average power constraint and satisfying a given effective covert rate requirement which we denote by $\tau$. In this section, we consider optimal choices for the parameters in our control to achieve the said purpose.

Although Alice's transmit power, $P_a$, is assumed to be fixed in this work, to make the problem feasible, we assume that the value of $P_a$ at least satisfies the rate requirement from Alice to Bob when no AN is transmitted by Bob. The rest of the design parameters that affect the performance of covert communication in our system model are the distribution parameters of Bob's AN power, $\{ P_{\text{min}}, P_{\text{max}} \} $, and the \textit{a priori} probabilities of Alice's transmission, $\{ \pi_0, \pi_1\} $, with $\pi_0 = 1 - \pi_1$.

We state our main problem as following:
\begin{equation}\label{p1}
\begin{aligned}
\textbf{P1} \quad  \underset{\pi_1, P_{\text{min}}, P_{\text{max}}}{\mathrm{maximize}} \quad &\overline{\mathbb{P}_{E}^*} \\
\mathrm{subject~to} \quad  &\pi_1 R_{ab} (1-\delta_{ab}) \geq \tau,  \\
&P_{\text{min}} + P_{\text{max}} \leq 2 P_{\text{avg}},
\end{aligned}
\end{equation}
while
\begin{equation}\label{ccc}
P_a \geq \frac{\lambda_{ab}\sigma_b^2 \left(2^{R_{ab}}-1\right)}{\ln \left(R_{ab}/\tau \right) }
\end{equation}
is assumed for feasibility. Here, the expression for $\overline{\mathbb{P}_{E}^*}$ is given in (\ref{4.8}) under Lemma $2$, $\delta_{ab}$ is the transmission outage probability and is a function of $\{ P_{\text{min}}, P_{\text{max}} \}$, $\tau$ is the minimum required covert rate, and $P_{\text{avg}}$ is the average transmit power for Bob's AN. We solve \textbf{P1} in a step-by-step manner, as this approach not only provides the globally optimal solution, but also provides further insights in the role of different parameters in achieving the said purpose of covertness.

\subsection{Optimal Minimum AN Power}
For a given average transmit power at Bob, we look to minimize the value of transmission outage probability, in order to satisfy the covert rate requirement, corresponding to the first constraint in (\ref{p1}), while maximizing $\overline{\mathbb{P}_{E}^*}$. Under these conditions, in this subsection, we consider finding the optimal minimum AN power at Bob, $P_{\text{min}}$, for any given maximum AN power, $P_{\text{max}}$, and prior probabilities of Alice's transmission, $\{\pi_0, \pi_1 \}$.

\begin{prop}
The optimal choice of $P_{\text{min}}$ to maximize the expected detection error probability at Willie, $\overline{\mathbb{P}_{E}^*}$, while satisfying the effective covert rate requirement from Alice to Bob is given by $P_{\text{min}}^*=0$.
\end{prop}

\begin{IEEEproof}
We first consider the maximization of $\overline{\mathbb{P}_{E}^*}$, where the optimal choice of $P_{\text{min}}$ should maximize $\kappa(t) = 1 + t \ln t - t^2$ under both cases of $\pi_1 \geq \pi_0$ and $\pi_1 < \pi_0$, as per (\ref{4.8}). To first determine the monotonicity of $\overline{\mathbb{P}_{E}^*}$ w.r.t. $P_{\text{min}}$, we consider the derivatives of $\kappa(t)$ w.r.t. $t$, given by $\frac{\partial \kappa(t)}{\partial t} = 1 + \ln t -2t $ and $\frac{\partial^2 \kappa(t)}{\partial t^2} = \frac{1}{t} - 2 $. Since $P_{\text{max}} \geq P_{\text{min}}$, we have

\begin{equation}\label{5.3}
\begin{aligned}
\frac{\partial t}{\partial P_{\text{min}}} &= \frac{\lambda_{aw}\lambda_{bw}P_a}{\left[ \lambda_{bw}P_a + \lambda_{aw}(P_{\text{max}}-P_{\text{min}}) \right]^2} \geq 0 .
\end{aligned}
\end{equation}

For $t \in [0,1)$, the first derivative of $\kappa(t)$ w.r.t. $t$ increases for $0 \leq t < 1/2$ and decreases for $1/2 \leq t <1$, with the maximum value of $-\ln 2$, occurring at $t=1/2$. Using this and the fact that $\frac{\partial t}{\partial P_{\text{min}}} \geq 0$, it can be concluded that $\kappa(t)$, and resultantly, $\overline{\mathbb{P}_{E}^*}$ is a decreasing function of $P_{\text{min}}$. Hence, the optimal choice in this regard is the minimum possible value of $P_{\text{min}}$, which is zero.

We next consider the covert rate constraint, where the outage probability $\delta_{ab}$ is represented as
\begin{equation}\label{5.4}
\delta_{ab} = 1 - \lambda_{bb} \exp (-\lambda_{ab}\mu \sigma_b^2) v(x),
\end{equation}
and here
\begin{equation}\label{5.5}
v(x) = \frac{1}{y-x} \ln \left( \frac{\lambda_{bb}+y}{\lambda_{bb}+x} \right),
\end{equation}
$x \triangleq \lambda_{ab}\phi \mu P_{\text{min}} \geq 0$, $y \triangleq \lambda_{ab}\phi \mu P_{\text{max}} \geq 0$, and $y \geq x$. Considering the first derivative of $v(x)$ w.r.t. $x$, we have
\begin{equation}\label{5.6}
\frac{\partial v(x)}{\partial x} = \frac{1}{\left(y-x \right)^2} \left[\ln \left(\frac{\lambda_{bb}+y}{\lambda_{bb}+x}\right) - \frac{y-x}{\lambda_{bb}+x} \right] .
\end{equation}
Here, $\frac{\partial v(x)}{\partial x}$ depends on
\begin{equation}\label{5.7}
\begin{aligned}
l(x) &\triangleq \ln \left(\frac{\lambda_{bb}+y}{\lambda_{bb}+x}\right) - \left( \frac{y-x}{\lambda_{bb}+x}\right) \\
&= \ln \left(1 + \frac{y-x}{\lambda_{bb}+x}\right) - \left(\frac{y-x}{\lambda_{bb}+x}\right) \leq 0 ,
\end{aligned}
\end{equation}
where the second line in (\ref{5.7}) is due to the logarithmic inequality, $\ln (1+a) \leq a, \forall a \geq -1$. Thus $v(x)$ is always a decreasing function of $x$, and resultantly, $\delta_{ab}$ is always an increasing function of $P_{\text{min}}$. From the covert rate constraint, we can write
 \begin{equation}\label{bbb}
 \delta_{ab} \leq 1 - \frac{\tau}{\pi_1 R_{ab}} ,
 \end{equation}
and hence, to satisfy this constraint, $P_{\text{min}}$ is upper bounded by a value which can be found by solving (\ref{bbb}) at equality. This concludes the proof.
\end{IEEEproof}

As a result of Proposition $1$, we can simplify the transmission outage probability at Bob and the expected detection error probability at Willie as
\begin{equation}\label{5.8}
\delta_{ab} = 1 - \frac{\lambda_{bb}\exp(- \lambda_{ab}\mu\sigma_b^2)}{P_{\text{max}} \lambda_{ab}\phi\mu } \ln \left[ \frac{\lambda_{bb}+\lambda_{ab}\phi\mu P_{\text{max}}}{\lambda_{bb}} \right],
\end{equation}
and
\begin{equation}\label{5.10}
\begin{aligned}
\overline{P_{E}^*} =
\begin{cases}
\pi_0 \left[ 1 + s \ln s - s^2    \right], & \text{if} \quad \pi_1 \geq \pi_0 \\
\pi_1 \left[ 1 + s \ln s - s^2    \right], & \text{otherwise} ,
\end{cases}
\end{aligned}
\end{equation}
respectively, where
\begin{equation}\label{def_s}
s =\frac{\lambda_{bw}P_a}{\lambda_{bw}P_a + \lambda_{aw}P_{\text{max}}} .
\end{equation}

\subsection{Optimal Priors for Alice's Transmission}
Once the optimal value of $P_{\text{min}}$ has been found, the task from Alice and Bob's perspective is to find the optimal \textit{a priori} probabilities of Alice's transmission and Bob's maximum possible transmit power, $P_{\text{max}}$. In this subsection, we consider finding the optimal choice of Alice's \textit{a priori} transmission probabilities for a given $P_{\text{max}}$. We state this problem as:
\begin{equation}\label{5.11}
\begin{aligned}
 \textbf{P1.1} \quad \underset{\pi_1}{\mathrm{maximize}} \quad &\overline{\mathbb{P}_{E}^*} \\
\mathrm{subject~to} \quad  &\pi_1 R_{ab} (1-\delta_{ab}) \geq \tau,  \\
\end{aligned}
\end{equation}
where the expression for $\overline{\mathbb{P}_{E}^*}$ is now given by (\ref{5.10}), and the feasibility condition of (\ref{ccc}) is still held. The solution to problem \textbf{P1.1} is presented in the following:

\begin{prop}
The optimal choice of \textit{a priori} probabilities for Alice's transmission, as a function of maximum AN power, $P_{\text{max}}$, is given by
\begin{equation}\label{5.12}
\pi_1^*(P_{\text{max}}) = \max \left(\frac{1}{2}, \frac{\tau}{R_{ab}(1-\delta_{ab}(P_{\text{max}}))} \right),
\end{equation}
and $\pi_0^* = 1 - \pi_1^*$.
\end{prop}

\begin{IEEEproof}
We consider the two cases for $\overline{\mathbb{P}_{E}^*}$ individually. Note here that $\delta_{ab}$ is now a function of $P_{\text{max}}$ only.

\subsubsection{$\pi_1 < \pi_0$}
In this case, $\pi_1 < 1/2$, and using the constraint in P1.1, we have $\frac{\tau}{R_{ab}(1-\delta_{ab}(P_{\text{max}}))} \leq \pi_1 < 1/2$, which can only happen when $\frac{\tau}{R_{ab}(1-\delta_{ab}(P_{\text{max}}))} \leq 1/2$. Also in this case, $\frac{\partial \overline{\mathbb{P}_{E}^*}}{\partial \pi_1} = 1 + s \ln s - s^2 \geq 0$ for $ s \in [0,1)$.

\subsubsection{$\pi_1 \geq \pi_0$}
Here, $\pi_1 \geq 1/2$, and due to the constraint in P1.1, $\pi_1 \geq \max \left(\frac{1}{2}, \frac{\tau}{R_{ab}(1-\delta_{ab}(P_{\text{max}}))}\right)$. Also, in this case, $\frac{\partial \overline{\mathbb{P}_{E}^*}}{\partial \pi_1} = -(1 + s \ln s - s^2) \leq 0$ for $s \in [0,1)$.

Combining these two cases gives the desired result.
\end{IEEEproof}

From Proposition $3$, it is evident that the optimal value of $\pi_1$ is dependent upon the choice of $P_{\text{max}}$. Thus to satisfy a given covert rate requirement, any choice of $P_{\text{max}}$ at Bob, directly affecting the transmission outage probability through self-interference, will determine whether $\pi_1^*$ is equal to $0.5$ or not. Since the purpose of our covert scheme is to maximize the detection error at Willie while satisfying the rate requirement, it presents an interesting interplay of our choice of these parameters.

\subsection{Optimal Maximum AN Power}
Once the optimal priors for Alice's transmission i.e., $\{\pi_0^*, \pi_1^* \}$ have been found in terms of $P_{\text{max}}$, the expected detection error probability at Willie becomes
\begin{equation}\label{5.14}
\begin{aligned}
\overline{\mathbb{P}_{E}^*}(\pi_1^*) =
\begin{cases}
\frac{1}{2} \kappa(s), & \text{if} \quad \frac{\tau}{R_{ab}(1 - \delta_{ab})} \leq \frac{1}{2} \\
\left(1 - \frac{\tau}{R_{ab}(1-\delta_{ab})} \right) \kappa(s), & \text{else} ,
\end{cases}
\end{aligned}
\end{equation}
where again, $\kappa(s) = \left(1 + s \ln s - s^2   \right)$, and $s$ is as defined earlier in (\ref{def_s}). We now consider finding the optimal value for Bob's maximum transmit power, $P_{\text{max}}$, under the average power constraint. This problem is stated as
\begin{equation}\label{5.15}
\begin{aligned}
\textbf{P1.2} \quad \underset{P_{\text{max}}}{\mathrm{maximize}} \quad &\overline{\mathbb{P}_{E}^*}(\pi_1^*) \\
\mathrm{subject~to} \quad  &\pi_1^* R_{ab} (1-\delta_{ab}) \geq \tau,  \\
&P_{\text{max}} \leq 2 P_{\text{avg}}.
\end{aligned}
\end{equation}
We note here that in the statement of \textbf{P1.2} above, $\overline{\mathbb{P}_{E}^*}$ from (\ref{5.10}) has now been replaced by $\overline{\mathbb{P}_{E}^*}(\pi_1^*)$ in (\ref{5.14}) and the feasibility condition of (\ref{ccc}) is still held. Following the step-by-step approach, and due to the monotonicity of $\overline{\mathbb{P}_E^*}$ w.r.t $P_{\text{min}}$ and $\pi_1$, \textbf{P1} is now reduced to \textbf{P1.2}. The solution to this problem is presented in the following proposition.

\begin{prop}
The optimal value for Bob's maximum transmit power under an average power constraint, $P_{\text{avg}}$, is given by
\begin{equation}\label{5.16}
\begin{aligned}
P_{\text{max}}^* =
\begin{cases}
2 P_{\text{avg}}, & \text{if} \quad 2 P_{\text{avg}} \leq P_{\text{max}}^{\dagger} \\
P_{\text{max}}^{\ddagger}, & \text{otherwise},
\end{cases}
\end{aligned}
\end{equation}
where $P_{\text{max}}^{\dagger}$ is the solution of $\delta_{ab}(P_{\text{max}}) = 1 - \frac{2\tau}{R_{ab}}$ for $P_{\text{max}}$ and $P_{\text{max}}^{\ddagger}$ is the solution to
\begin{equation}\label{5.17}
\begin{aligned}
\underset{P_{\text{max}}}{\mathrm{maximize}} \quad &\left(1 - \frac{\tau}{R_{ab}(1-\delta_{ab}(P_{\text{max}}))} \right) \left(1 + s \ln s - s^2   \right) \\
\mathrm{subject~to} \quad &P_{\text{max}}^{\dagger} \leq P_{\text{max}} \leq 2 P_{\text{avg}} .
\end{aligned}
\end{equation}
\end{prop}

\begin{IEEEproof}
We first show the monotonicity of $\delta_{ab}$ w.r.t $P_{\text{max}}$. Here, $\delta_{ab}$ can be written as
\begin{equation}\label{5.18}
\delta_{ab} = 1 - \lambda_{bb} \exp (- \lambda_{ab}\mu \sigma_b^2) u(x) ,
\end{equation}
where $u(x) \triangleq \frac{1}{x} \ln \left( \frac{\lambda_{bb}+x}{\lambda_{bb}}\right)$ and $x \triangleq \lambda_{ab}\phi\mu P_{\text{max}} \geq 0$. We note that
\begin{equation}
\frac{\partial u(x)}{\partial x} = \frac{1}{x^2}\left( \frac{x}{\lambda_{bb}+x} - \ln \left( \frac{\lambda_{bb}+x}{\lambda_{bb}} \right)  \right),
\end{equation}
which depends on $m(x) \triangleq \frac{x}{\lambda_{bb}+x} - \ln \left( \frac{\lambda_{bb}+x}{\lambda_{bb}} \right)$. Here, $m(0)=0$ and $\frac{\partial m(x)}{\partial x} = - \frac{x}{(\lambda_{bb}+x)^2} \leq 0$, thus $m(x)$ decreases monotonically with $x$, giving $m(x) \leq m(0)$ for $x \geq 0$, and resultantly, $\frac{\partial u(x)}{\partial x} \leq 0$. As a result, $\delta_{ab}$ is a monotonically increasing function of $P_{\text{max}}$.

Next, we consider the optimal choice of $P_{\text{max}}$ under the two cases of $\overline{\mathbb{P}_E^*}$, keeping in view the change in $\delta_{ab}$ w.r.t. $P_{\text{max}}$.

\setcounter{subsubsection}{0}
\subsubsection{}
For $\frac{\tau}{R_{ab}(1-\delta_{ab})} \leq \frac{1}{2}$, we have
\begin{equation}\label{5.19}
\delta_{ab} \leq 1 - \frac{2\tau}{R_{ab}}.
\end{equation}
Due to a monotonic increase in $\delta_{ab}$ w.r.t. $P_{\text{max}}$, the optimal value of $P_{\text{max}}$ has to satisfy $P_{\text{max}} \leq P_{\text{max}}^{\dagger}$, where $P_{\text{max}}^{\dagger}$ is the solution of (\ref{5.19}) at equality. Combining with the average power constraint, we have $P_{\text{max}} \leq \min \left(2 P_{\text{avg}}, P_{\text{max}}^{\dagger} \right)$. Now we consider the monotonicity of $\overline{\mathbb{P}_{E}^*}(\pi_1^*) = \frac{1}{2}(1+s \ln s -s^2)$ w.r.t $P_{\text{max}}$. Here, $ \frac{\partial \overline{\mathbb{P}_{E}^*}(\pi_1^*)}{\partial s} = \frac{1}{2} \left(1 + \ln s - 2s \right)$ and $\frac{\partial \overline{\mathbb{P}_{E}^*}(\pi_1^*)^2}{\partial^2 s} = \frac{1}{2s} - 1 $, where $s = \frac{\lambda_{bw}P_a}{\lambda_{bw}P_a + \lambda_{aw}P_{\text{max}}}$. Also,

\begin{equation}
\frac{\partial s}{\partial P_{\text{max}}} = - \frac{\lambda_{aw}\lambda_{bw}P_a}{\left(\lambda_{bw}P_a + \lambda_{aw}P_{\text{max}} \right)^2} \leq 0 .
\end{equation}
We note here that $s \in [0,1)$, $\frac{\partial \overline{\mathbb{P}_{E}^*}(\pi_1^*)}{\partial s}$ increases for $0 \leq s <1/2$ and decreases for $1/2 \leq s <1$, with a maximum value of $- \frac{1}{2}\ln 2$. Since $\frac{\partial s}{\partial P_{\text{max}}} \leq 0$, $\overline{\mathbb{P}_{E}^*}(\pi_1^*)$ is an increasing function of $P_{\text{max}}$, and hence the best possible choice in this case is $P_{\text{max}} = \min \left(2 P_{\text{avg}}, P_{\text{max}}^{\dagger} \right)$.
\subsubsection{}
For $\frac{\tau}{R_{ab}(1-\delta_{ab})} > \frac{1}{2}$, we have
\begin{equation}\label{5.20}
\delta_{ab} > 1 - \frac{2\tau}{R_{ab}},
\end{equation}
and in this case,
\begin{equation}
\overline{\mathbb{P}_{E}^*}(\pi_1^*) = \left(1 - \frac{\tau}{R_{ab}(1-\delta_{ab})} \right) \left(1 + s \ln s - s^2   \right).
\end{equation}
Since $\delta_{ab}$ increases monotonically in $P_{\text{max}}$, hence to satisfy (\ref{5.20}), $P_{\text{max}} > P_{\text{max}}^{\dagger}$, and resultantly, the optimal choice lies between $P_{\text{max}}^{\dagger}$ and $2 P_{\text{avg}}$, where $P_{\text{max}}^{\dagger}$ is as defined earlier. Let $\overline{\mathbb{P}_{E}^*}(\pi_1^*) = p(x)q(x) $, where $p(P_{\text{max}})=\left(1 - \frac{\tau}{R_{ab}(1-\delta_{ab}(P_{\text{max}}))} \right)$ and $q(P_{\text{max}})=\left(1 + s \ln s - s^2   \right)$. We note here that $\overline{\mathbb{P}_{E}*}(\pi_1^*)$ is not a monotonic function of $P_{\text{max}}$, since as $P_{\text{max}}$ increases, $p(P_{\text{max}})$ decreases while $q(P_{\text{max}})$ increases. Thus there may exist an optimal value of $P_{\text{max}}$ that maximizes $\overline{\mathbb{P}_{E}^*}(\pi_1^*)$, which motivates the optimization
\begin{equation}\label{5.21}
P_{\text{max}}^{\ddagger} = \underset{P_{\text{max}}}{\mathrm{maximize}} \quad \overline{\mathbb{P}_{E}^*}(\pi_1^*).
\end{equation}
We note that the optimization problem in (\ref{5.21}) is of one dimension and can be solved by methods of efficient numerical search.

Combining the two cases, the optimal value for $P_{\text{max}}$ is found, thus completing the proof.
\end{IEEEproof}

\begin{remark}
The approach we have taken in solving $\mathbf{P1}$ guarantees that the obtained solution is globally optimal. Specifically, we first solved for the optimal $P_{\text{min}}$ for any value of $\pi_1$ and $P_{\text{max}}$. We next solved for the optimal $\pi_1$ as a function of any given $P_{\text{max}}$. Finally the optimal $P_{\text{max}}$ is obtained. The globally optimal solution to $\mathbf{P1}$ can thus be summarized as $P_{\text{min}}^* = 0$, $\pi_1^* = \max \left(1/2, \tau/[R_{ab}(1-\delta_{ab}(P_{\text{max}}^*))] \right)$ and $P_{\text{max}}^*$ as given in $(\ref{5.16})-(\ref{5.17})$.
\end{remark}

As discussed in Remark $2$, increasing the value of $P_{\text{max}}$ helps improve our covert performance, but on the other hand, $P_{\text{max}}$ also affects the covertly conveyed information through self-interference. Proposition $4$ tells us that while satisfying the average power constraint, the optimal choice of $P_{\text{max}}$ satisfies the covert rate requirement with a bare equality, while maximizes the expected detection error probability at Willie. There might exist values of $\tau$ for which the choice of $P_{\text{max}}$ satisfying the rate constraint under $\pi=\frac{1}{2}$ are not the optimal choice of maximizing $\overline{P_E^*}$. Under such a scenario, the freedom of adjusting $\pi_1$ helps us in meeting the rate requirement while keeping $\overline{P_E^*}$ as high as possible.

\section{Numerical Results and Discussion}

\begin{figure}[t]
\centering
	\includegraphics[width=\linewidth]{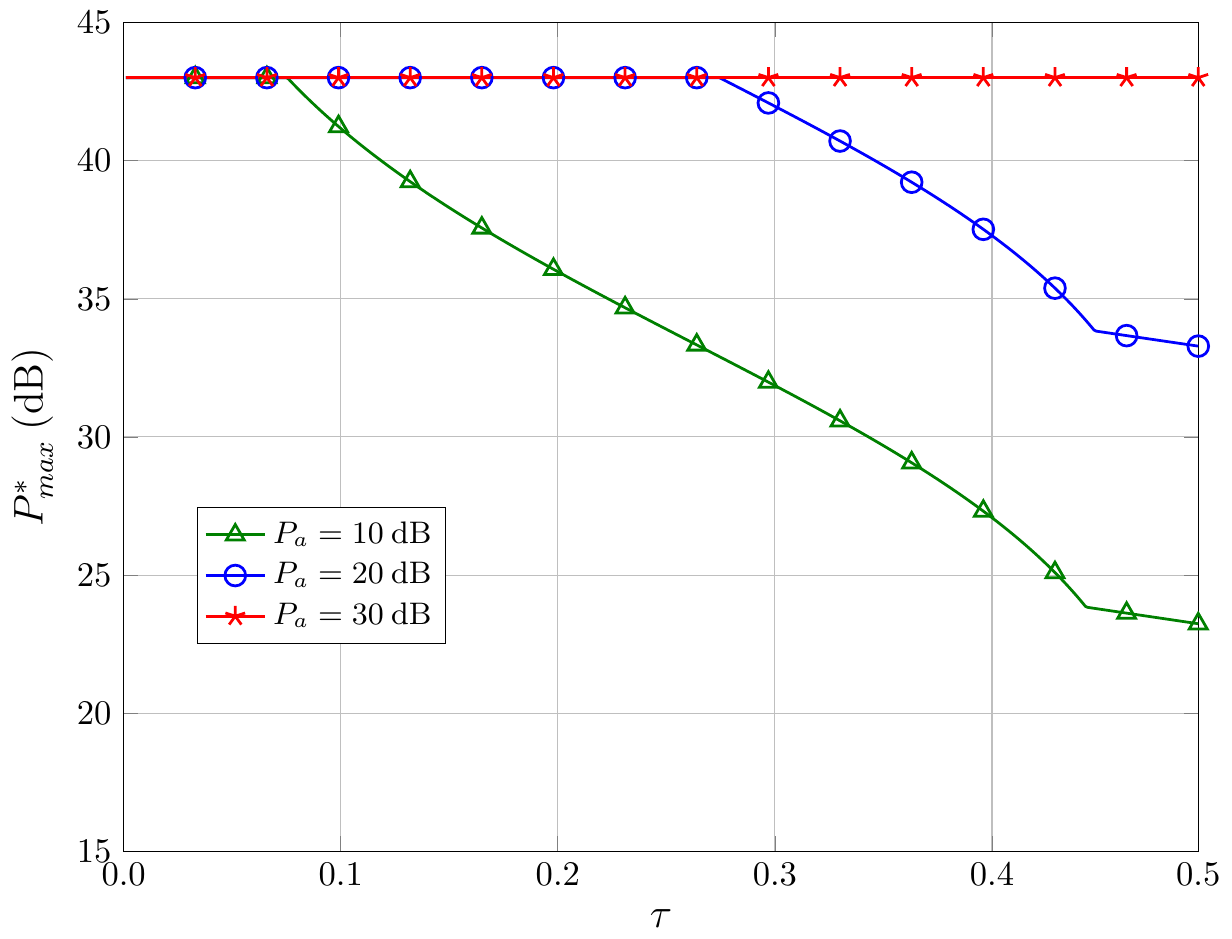}
	\caption{Optimal maximum transmit power of Bob's AN, $P_{max}^*$, versus the covert rate requirement from Alice to Bob, $\tau$, for varying values of Alice's transmit power, $P_a$.}
	\label{fig_sim_1}
\end{figure}

\begin{figure}[t]
\centering
	\includegraphics[width=\linewidth]{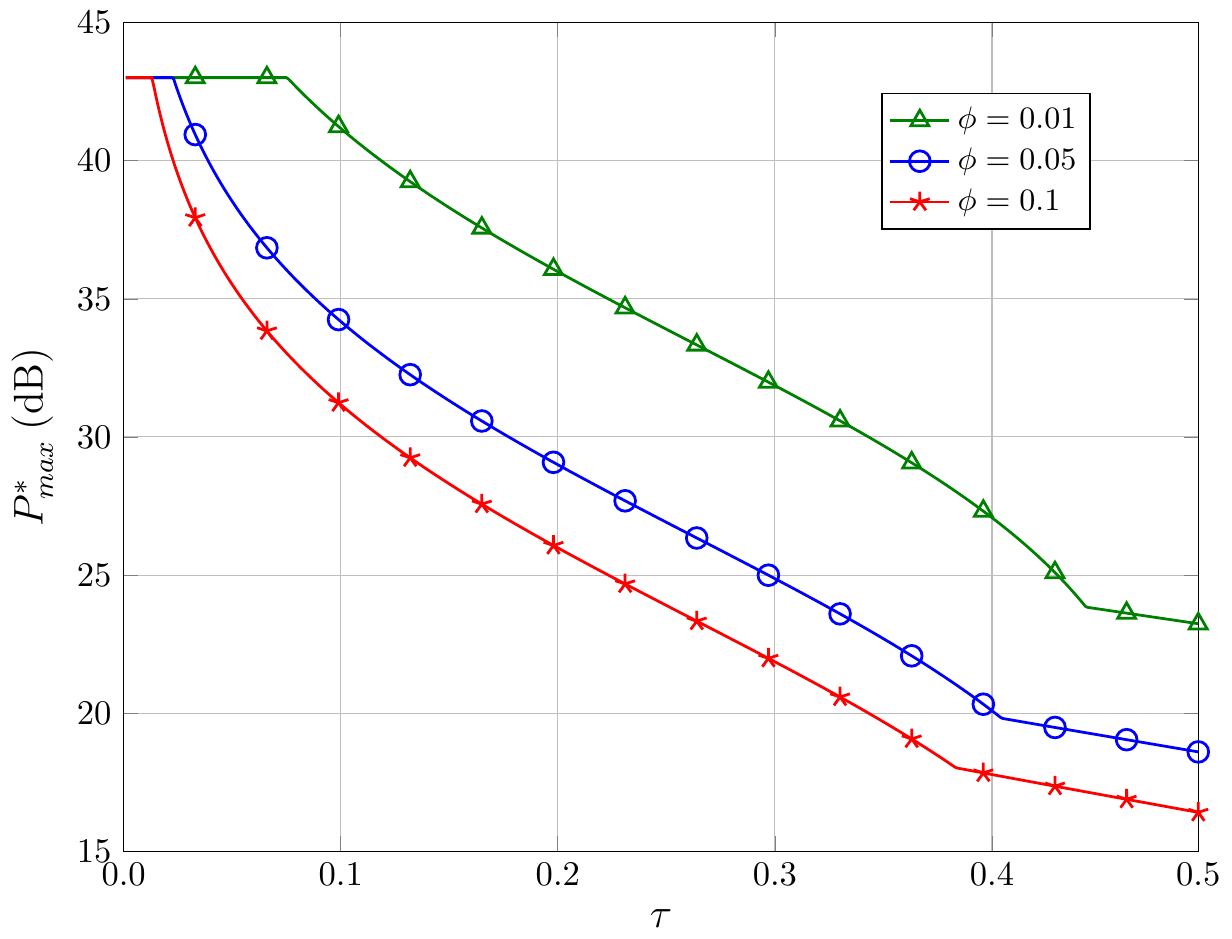}
	\caption{Optimal maximum transmit power of Bob's AN, $P_{max}^*$, versus the covert rate requirement from Alice to Bob, $\tau$, for varying values of Bob's self-interference cancellation coefficient, $\phi$.}
	\label{fig_sim_2}
\end{figure}

In this section, we present the numerical results and study the performance of our proposed scheme in achieving covertness while satisfying a given covert rate requirement. Unless otherwise stated, we set the transmit power at Alice $P_a = 10\:\text{dB}$, a pre-determined rate for Alice to Bob transmission $R_{ab}=1$, Bob and Willie's receiver noise power $\sigma_{b}^{2} = \sigma_{w}^{2} = -10\:\text{dB}$ and Bob's self-interference cancellation coefficient\footnote{Self-interference passive suppression of roughly $34 - 44$ dB for FD systems has been reported in the literature \cite{duarte_fd_2012}, while a combination of passive suppression and active cancellation resulting in a total self-interference suppression of $90$ dB has also been demonstrated \cite{passive_fd_2014}.} $\phi=0.01$.The average power constraint on Bob's AN power is $40\:\text{dB}$, while for simplicity, the means of all fading channels are considered as $1 / \lambda_{ab}=1 / \lambda_{aw}=1 / \lambda_{bw}=1 / \lambda_{bb}=1$.

We first show the effect of $P_a$ and $\phi$ on the optimal maximum transmit power for Bob's AN for varying covert transmission rate requirements, as demonstrated in Fig. \ref{fig_sim_1} and Fig. \ref{fig_sim_2}, respectively. In Fig. \ref{fig_sim_1} with a fixed value of $\phi$, a higher value of $P_a$ allows a higher value of $P_{max}^*$ to maximize the detection error probability at Willie, whilst satisfying the given rate requirement. In Fig. \ref{fig_sim_2}, with a fixed value of $P_a$ in the feasible range, a higher value of $\phi$ (i.e., poorer self-interference cancellation) requires a lower value of $P_{max}^*$ (i.e., less self-interference) to satisfy the same rate requirement. We note here that in such circumstances, a reduced $P_{max}^*$ for a given $P_a$ will adversely affect the achievable covertness.

\begin{figure}[t]
\centering
	\includegraphics[width=\linewidth]{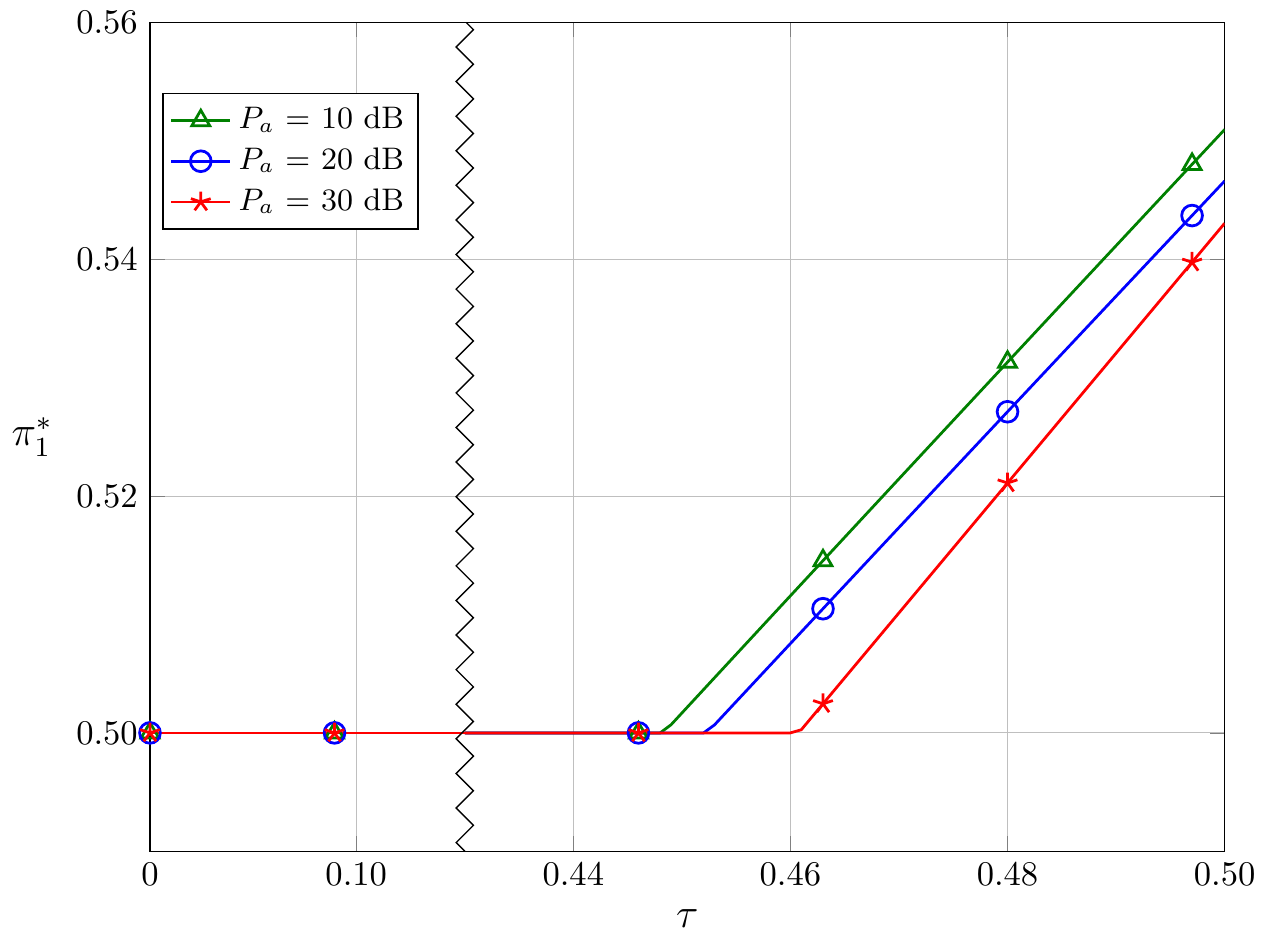}
	\caption{Optimal choice of transmission probability, $\pi_1^*$, versus the covert rate requirement from Alice to Bob, $\tau$, for varying values of Alice's transmit power, $P_a$.}
	\label{fig_sim_3}
\end{figure}

\begin{figure}[t!]
\centering
	\includegraphics[width=\linewidth]{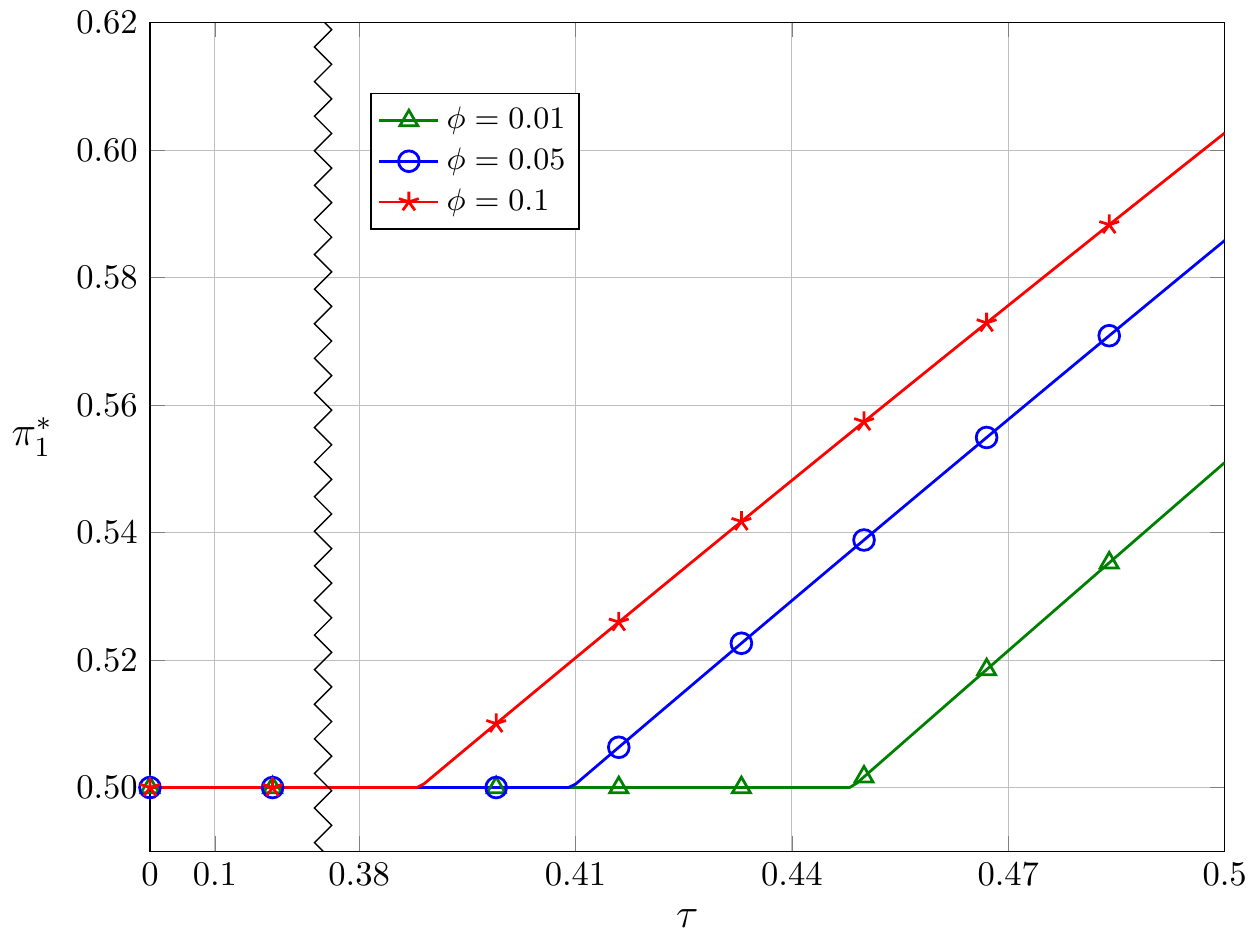}
	\caption{Optimal choice of transmission probability, $\pi_1^*$, versus the covert rate requirement from Alice to Bob, $\tau$, for varying values of Bob's self-interference cancellation coefficient, $\phi$. }
	\label{fig_sim_4}
\end{figure}

We next consider the effect of $P_a$ and $\phi$ on the optimal transmission probability of Alice's covert transmissions for varying covert transmission rate requirements, as demonstrated in Fig. \ref{fig_sim_3} and Fig. \ref{fig_sim_4}, respectively. From Fig. \ref{fig_sim_3}, we see that for a given $P_a$, a choice of $\pi_1 = 1/2$ is optimal up to a certain value of $\tau$, but a further increase in $\tau$ results in an increase in optimal $\pi_1$. For a given $P_a$, a rate requirement can be met by decreasing the value of $P_{max}^*$, but it will in return decrease the achievable covertness. Keeping in view the results shown in Fig. \ref{fig_sim_1} and Fig. \ref{fig_sim_2}, the optimal solution dictates that instead of making a drastic change in $P_{max}^*$, a better choice is to decrease $P_{max}^*$ a little while $\pi_1$ can be increased to meet the rate requirement.

\begin{figure}[t]
\centering
	\includegraphics[width=\linewidth]{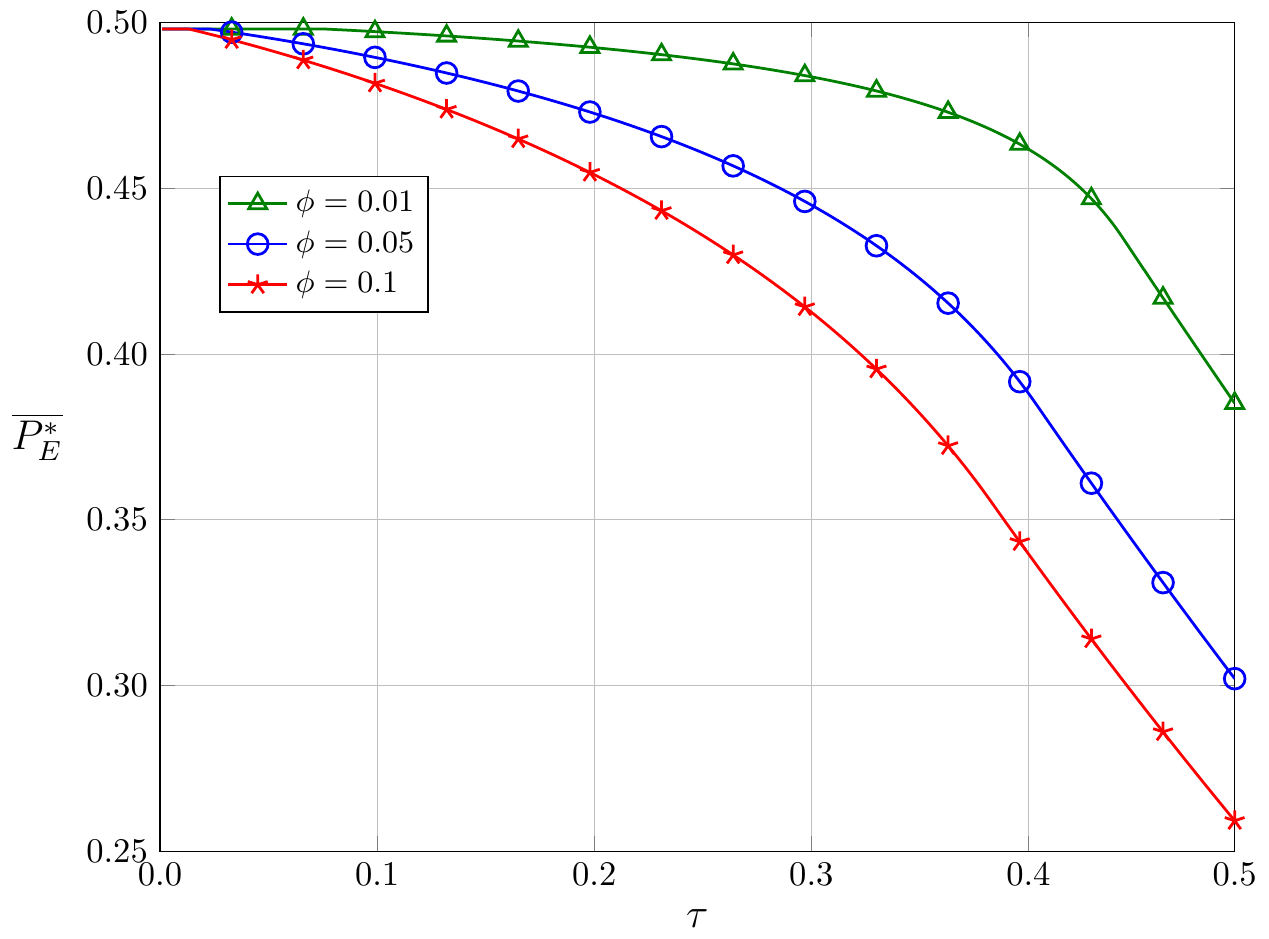}
	\caption{The expected detection error probability at Willie, $\overline{P_E^*}$, versus the covert rate requirement from Alice to Bob, $\tau$, for varying values of Bob's self-interference cancellation coefficient, $\phi$. }
	\label{fig_sim_5}
\end{figure}

\begin{figure}[h!]
\centering
	\includegraphics[width=\linewidth]{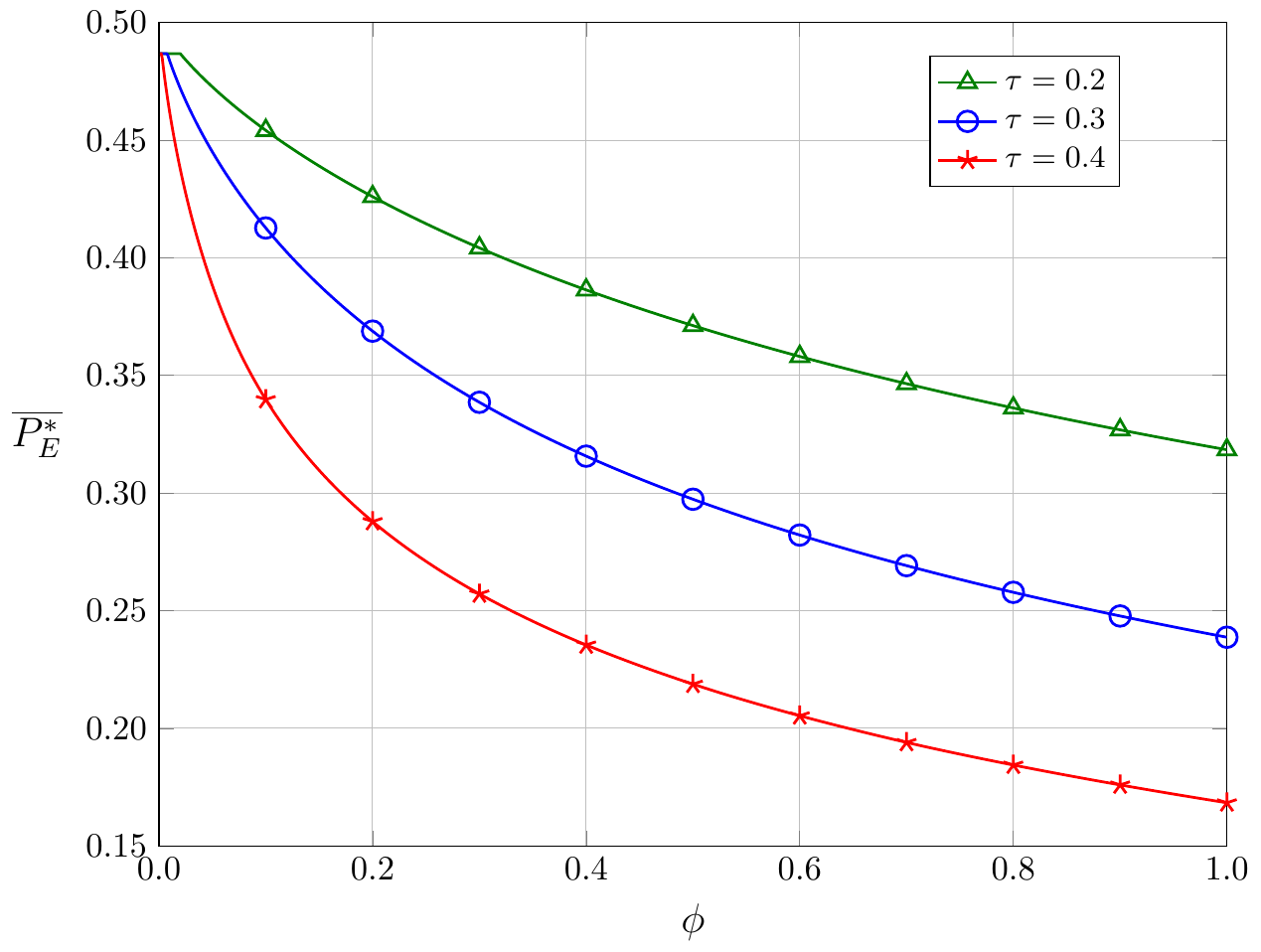}
	\caption{The expected minimum detection error probability at Willie, $\overline{P_E^*}$, versus Bob's self-interference coefficient $\phi$, for varying values of covert rate requirement $\tau$.}
	\label{fig_6}
\end{figure}

As the value of $P_a$ is increased, the same effect appears for a little higher value of $\tau$. Fig. \ref{fig_sim_4} shows the effect of increasing $\phi$ on the optimal $\pi_1$ for a given value of $P_a$, which is inverse of what is observed for increasing $P_a$. Since an increase in $\phi$ will have a detrimental effect on the transfer of covert information, thus to keep the covertness high and to satisfy the rate requirement, an increase in $\pi_1$ is desired for an even lower value of $\tau$. To further demonstrate the effect of $\phi$, we show the expected detection error probability at Willie for different values of $\phi$ in Fig. \ref{fig_sim_5}. For a fixed $P_a$ and a given $\phi$, as $\tau$ increases, $P_{max}$ at Bob has to decrease in order to reduce Bob's self-interference. A lower value of $P_{\text{max}}$ will result in a lower $\overline{\mathbb{P}_E^*}$, since it decreases the confusion in received signal statistics at Willie. This effect is shown more explicitly in Fig. 8, where we show the effect of $\phi$  on the performance of proposed covert scheme through the expected minimum detection error probability at Willie $\overline{P_E^*}$. It should be noted here that a value of $\phi=0$ corresponds to a perfect cancellation of the self-interference while $\phi=1$ refers to no cancellation or suppression at all, representing the worst case scenario for the FD receiver. For a higher value of $\phi$, Bob has to reduce $P_{max}^*$ to satisfy a certain rate requirement, which in effect, reduces the achievable covertness.

\begin{figure}[t]
\centering
	\includegraphics[width=\linewidth]{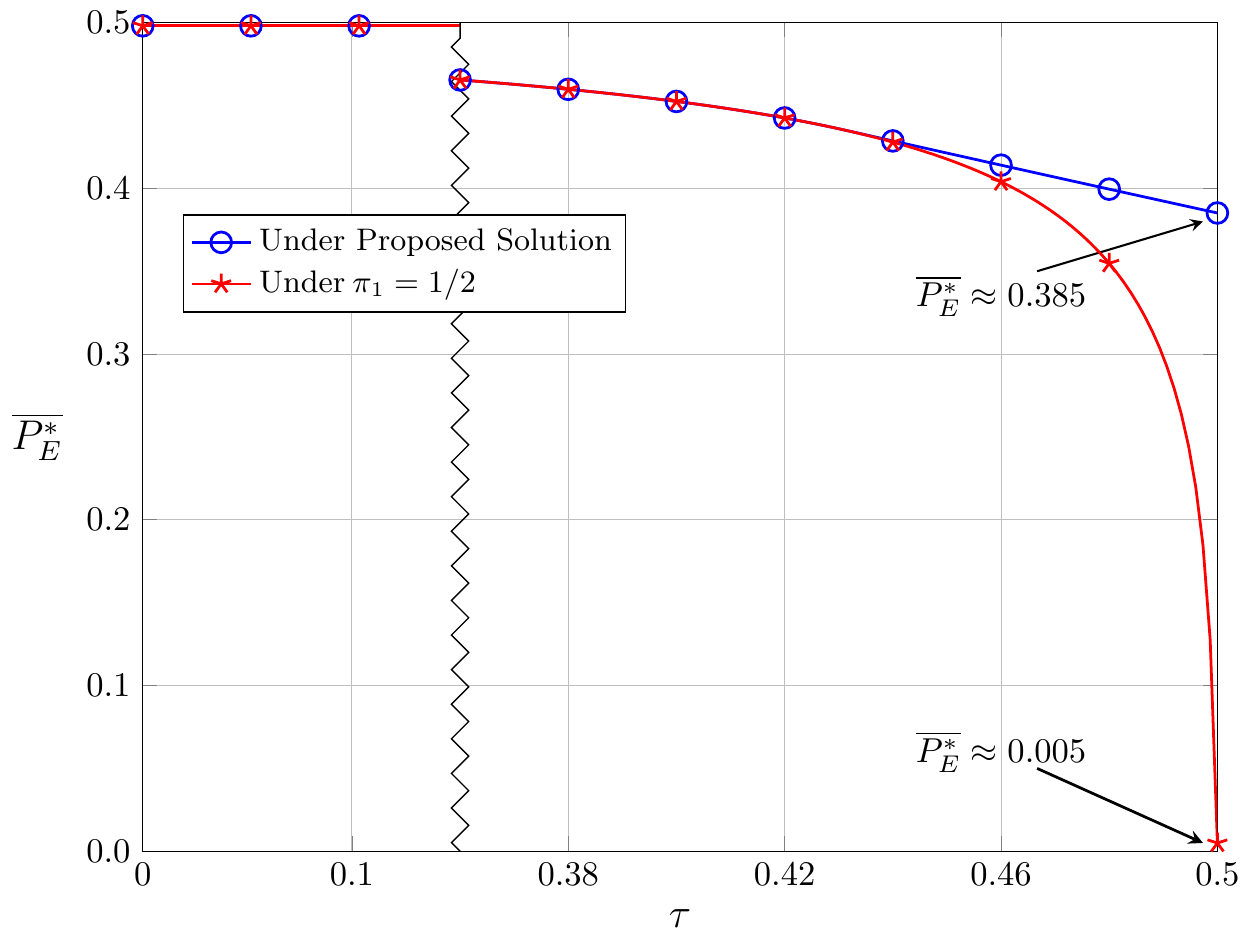}
	\caption{The expected detection error probability at Willie, $\overline{P_E^*}$, versus the covert rate requirement from Alice to Bob, $\tau$, under the proposed scheme and under the approach where $\pi_1 = 1/2$. }
	\label{fig_sim_7}
\end{figure}

Last but not least, we investigate the advantage of our proposed scheme of jointly optimizing $\pi_1$ and $P_{\text{max}}$ over a benchmark scheme of only optimizing $P_{\text{max}}$ while keeping $\pi_1 = 0.5$. Fig. \ref{fig_sim_7} shows the overall performance of our proposed scheme in terms of the expected detection error probability at Willie versus the covert rate requirement from Alice to Bob. For $\tau \in [0, 0.44]$, the proposed joint optimization scheme performs the same as the benchmark scheme and there is no discernable difference in $\overline{P_E^*}$ for the two schemes. However, for $\tau \geq 0.44$, the optimal $\pi_1$ starts to deviate from $0.5$, as shown in Fig. \ref{fig_sim_3} and Fig. \ref{fig_sim_4}. Here, the $\overline{P_E^*}$ achieved by the joint optimization scheme reduces gradually as the rate requirement increases, but the $\overline{P_E^*}$ for the benchmark scheme drops sharply, and at $\tau=0.5$, the benchmark scheme offers $\overline{P_E^*} \approx 0.005$, which means almost no covertness at all. Thus for $\tau \geq 0.44$, the proposed joint optimization scheme provides a significant gain in the achievable covertness.

\section{Conclusion}
In this paper, we have considered the potential of achieving covert communication using a full-duplex receiver that generates artificial noise to cause detection errors at a watchful adversary Willie. Considering a radiometer as the detector of choice at Willie, we have analyzed the conditions under which Willie makes detection errors, and characterized Willie's optimal detection performance conditioned over the fading channel realizations. From the perspective of cover communication pair, we have provided design guidelines for the optimal choice of transmit power of full-duplex receiver's artificial noise. Owing to the self-interference of the full-duplex receiver, these power levels need to be controlled carefully, otherwise they affect the transfer of any covert information. We have also shown that contrary to a commonly adopted assumption, the \textit{a priori} transmission probabilities of $0.5$ are not always the optimal choice to achieve the best possible covertness. A limitation of the current work lies in its assumption of $n \rightarrow \infty$. Future work will focus on scenarios where $n$ is finite. This will provide more precise results relating to future communication systems, especially for delay-intolerant systems.

\section*{Appendix A \\ Proof of Proposition $1$}

\begin{figure}[t]
\centering
	\includegraphics[width=\linewidth]{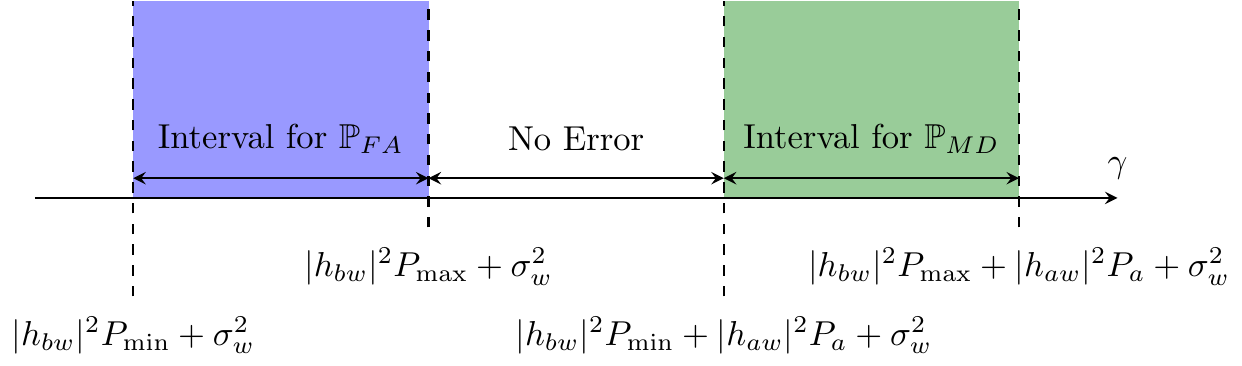}
	\caption{Case-I : $ |h_{bw}|^2 P_{\text{max}}+\sigma_w^2 < |h_{bw}|^2 P_{\text{min}} + |h_{aw}|^2 P_a+\sigma_w^2$}
	\label{fig_case_1}
\end{figure}

Using the definition of incorrect decisions at Willie, we have
\begin{equation}\label{3.3}
  \begin{aligned}
  \mathbb{P}_{FA} &= \mathcal{P}\left[D_1 | H_0  \right] = \mathcal{P}\left[P_w > \gamma^{} | H_0 \right] \\
  &= \mathcal{P}\left[  |h_{bw}|^2 P_b + \sigma_{w}^{2} > \gamma^{} \right] =  \mathcal{P} \left[P_b > \frac{\gamma^{}-\sigma_{w}^{2}}{|h_{bw}|^2}   \right] \\
  &= \begin{cases}
1, & \text{if} \quad \frac{\gamma^{} - \sigma_w^2}{|h_{bw}|^2} \leq P_{\text{min}} \\
\frac{|h_{bw}|^2 P_{\text{max}} + \sigma_w^2 - \gamma^{}}{|h_{bw}|^2 \left(P_{\text{max}}-P_{\text{min}}\right)},  & \text{if} \quad P_{\text{min}} < \frac{\gamma^{} - \sigma_w^2}{|h_{bw}|^2} \leq P_{\text{max}} \\
0, & \text{else},
\end{cases}
  \end{aligned}
  \end{equation}
 and
\begin{equation}\label{3.4}
 \begin{aligned}
  \mathbb{P}_{MD} &= \mathcal{P}\left[D_0 | H_1 \right] = \mathcal{P}\left[P_w < \gamma^{} | H_1 \right] \\
  &= \mathcal{P}\left[ |h_{bw}|^2 P_b + |h_{aw}|^2 P_{a} + \sigma_{w}^{2} < \gamma^{}  \right]   \\
  &=  \mathcal{P} \left[P_b < \frac{\gamma^{} - |h_{aw}|^2 P_{a}-\sigma_w^2}{|h_{bw}|^2}   \right] \\
 &= \begin{cases}
0, & \text{if} \quad  \nu \leq P_{\text{min}} \\
\frac{\gamma^{} -|h_{aw}|^2 P_a - |h_{bw}|^2 P_{\text{min}} -\sigma_w^2}{|h_{bw}|^2 \left(P_{\text{max}} - P_{\text{min}}\right)},  & \text{if} \quad P_{\text{min}} < \nu \leq P_{\text{max}} \\
1, & \text{else}.
\end{cases}
  \end{aligned}
  \end{equation}
where  $\nu \triangleq \frac{\gamma^{} - \sigma_w^2 -|h_{aw}|^2 P_a}{|h_{bw}|^2}$, $h_{aw}$ and $h_{bw}$ denote the channels from Alice and Bob to Willie, respectively, $\mathcal{P}[\cdot]$ denotes the probability measure and we have used the conditioning over the uniform distribution of Bob's transmit power, i.e.,  $P_b \sim \mathcal{U}(P_{\text{min}}, P_{\text{max}})$. Since Willie has to choose the threshold of his detector, $\gamma$, such that the probability of error at Willie, $\mathbb{P}_{E}=\pi_0\mathbb{P}_{FA} + \pi_1\mathbb{P}_{MD}$, is minimized, thus Willie considers the following:
\begin{equation} \label{3.11}
 \underset{\gamma}{\mathrm{minimize}} \quad \pi_0\mathbb{P}_{FA} + \pi_1\mathbb{P}_{MD},
\end{equation}
where the expressions for incorrect decisions for individual slots at Willie are as defined earlier in (\ref{3.3}) and (\ref{3.4}). Willie chooses his detector's threshold, in the intervals marked by the quantities given by $|h_{bw}|^2 P_{\text{min}}+\sigma_w^2$, $|h_{bw}|^2 P_{\text{min}} + |h_{aw}|^2 P_a+\sigma_w^2$, $|h_{bw}|^2 P_{\text{max}}+\sigma_w^2$ and $|h_{bw}|^2 P_{\text{max}}+ |h_{aw}|^2 P_a+\sigma_w^2 $. We also note that
\begin{itemize}
\item $|h_{bw}|^2 P_{\text{min}}+\sigma_w^2 \leq |h_{bw}|^2 P_{\text{max}}+ |h_{aw}|^2 P_a+\sigma_w^2$, but the relationship between $|h_{bw}|^2 P_{\text{min}} + |h_{aw}|^2 P_a+\sigma_w^2$ and $|h_{bw}|^2 P_{\text{max}}+\sigma_w^2$ is unclear.
\item For a choice of $\gamma < |h_{bw}|^2 P_{\text{min}}+\sigma_w^2$, $\mathbb{P}_{FA}=1$, $\mathbb{P}_{MD}=0$ and hence $\mathbb{P}_{E}=\pi_0$.
\item For a choice of $\gamma > |h_{bw}|^2 P_{\text{max}}+ |h_{aw}|^2 P_a+\sigma_w^2$, $\mathbb{P}_{FA}=0$, $\mathbb{P}_{MD}=1$ and hence $\mathbb{P}_{E}=\pi_1$.
\end{itemize}

In the following, we analyse the error probability at Willie under the two different cases of $|h_{bw}|^2 P_{\text{min}} + |h_{aw}|^2 P_a+\sigma_w^2 \leq |h_{bw}|^2 P_{\text{max}}+\sigma_w^2$ and $|h_{bw}|^2 P_{\text{min}} + |h_{aw}|^2 P_a+\sigma_w^2 > |h_{bw}|^2 P_{\text{max}}+\sigma_w^2$.

\begin{figure}[t]
\centering
	\includegraphics[width=\linewidth]{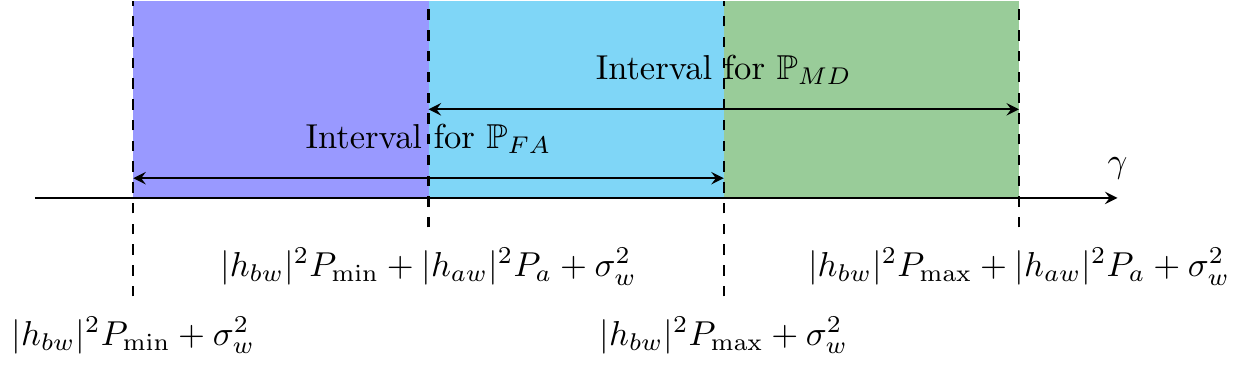}
	\caption{Case-II : $ |h_{bw}|^2 P_{\text{max}}+\sigma_w^2 \geq |h_{bw}|^2 P_{\text{min}} + |h_{aw}|^2 P_a+\sigma_w^2$}
	\label{fig_case_2}
\end{figure}

\subsection*{Case - I : $ |h_{bw}|^2 P_{\text{max}}+\sigma_w^2 < |h_{bw}|^2 P_{\text{min}} + |h_{aw}|^2 P_\text{a}+\sigma_w^2$}
This case is graphically shown in Fig. \ref{fig_case_1}  and we have three intervals for the choice of $\gamma$.

\subsubsection*{(1) $|h_{bw}|^2 P_{\text{min}}+\sigma_w^2< \gamma \leq |h_{bw}|^2 P_{\text{max}}+\sigma_w^2$}
In this case, $\mathbb{P}_{MD}=0$, and
\begin{equation}\label{3.12}
\mathbb{P}_{E} = \pi_0 \mathbb{P}_{FA} = \pi_0 \left[ \frac{|h_{bw}|^2 P_{\text{max}}+\sigma_w^2 -\gamma}{|h_{bw}|^2 \left(P_{\text{max}} - P_{\text{min}}  \right) } \right],
\end{equation}
which has a decreasing partial derivative with respect to (w.r.t.) $\gamma$, given by $\frac{-\pi_0}{|h_{bw}|^2 \left(P_{\text{max}} - P_{\text{min}} \right)}$, thus $\gamma = |h_{bw}|^2 P_{\text{max}}+\sigma_w^2$ should be chosen.

\subsubsection*{(2) $|h_{bw}|^2 P_{\text{min}} + |h_{aw}|^2 P_\text{a}+\sigma_w^2 < \gamma \leq |h_{bw}|^2 P_{\text{max}}+ |h_{aw}|^2 P_\text{a}+\sigma_w^2$}
In this case, $\mathbb{P}_{FA}=0$, and
\begin{equation}\label{3.13}
\begin{aligned}
\mathbb{P}_{E} = \pi_1 \mathbb{P}_{MD} = \pi_1 \left[ \frac{\gamma - |h_{aw}|^2 P_\text{a} - |h_{bw}|^2 P_{\text{min}} -\sigma_w^2}{|h_{bw}|^2 \left(P_{\text{max}} - P_{\text{min}}  \right) } \right],
\end{aligned}
\end{equation}
which has an increasing partial derivative w.r.t $\gamma$, given by $\frac{\pi_1}{|h_{bw}|^2 \left(P_{\text{max}} - P_{\text{min}} \right)}$, thus $\gamma = |h_{bw}|^2 P_{\text{min}} + |h_{aw}|^2 P_\text{a}+\sigma_w^2$ should be chosen.

\subsubsection*{(3) $|h_{bw}|^2 P_{\text{max}}+\sigma_w^2 < \gamma \leq |h_{bw}|^2 P_{\text{min}} + |h_{aw}|^2 P_\text{a}+\sigma_w^2$}
In this case, $\mathbb{P}_{FA}=0$ and $\mathbb{P}_{MD}=0$, which means that a choice of $\gamma$ in this interval will have no detection errors at Willie.

\subsection*{Case - II : $ |h_{bw}|^2 P_{\text{max}}+\sigma_w^2 \geq |h_{bw}|^2 P_{\text{min}} + |h_{aw}|^2 P_\text{a}+\sigma_w^2$}
This case is graphically shown in Fig. \ref{fig_case_2} and we have three intervals for the choice of $\gamma$.

\subsubsection*{(1) $ |h_{bw}|^2 P_{\text{min}}+\sigma_w^2 < \gamma \leq |h_{bw}|^2 P_{\text{min}} + |h_{aw}|^2 P_\text{a}+\sigma_w^2 $}
In this case, $\mathbb{P}_{MD}=0$ and
\begin{equation}\label{3.14}
\mathbb{P}_{E} = \pi_0 \mathbb{P}_{FA} = \pi_0 \left[ \frac{|h_{bw}|^2 P_{\text{max}}+\sigma_w^2 -\gamma}{|h_{bw}|^2 \left(P_{\text{max}} - P_{\text{min}}  \right) } \right] ,
\end{equation}
which has a decreasing partial derivative w.r.t $\gamma$, given by $\frac{-\pi_0}{|h_{bw}|^2 \left(P_{\text{max}} - P_{\text{min}}\right)}$, thus $\gamma = |h_{bw}|^2 P_{\text{min}}+|h_{aw}|^2 P_\text{a}+\sigma_w^2$ should be chosen.

\subsubsection*{(2) $|h_{bw}|^2 P_{\text{max}}+\sigma_w^2 < \gamma \leq |h_{bw}|^2 P_{\text{max}}+ |h_{aw}|^2 P_\text{a}+\sigma_w^2$}
In this case, $\mathbb{P}_{FA}=0$, and
\begin{equation}\label{3.15}
\begin{aligned}
\mathbb{P}_{E} = \pi_1 \mathbb{P}_{MD} = \pi_1 \left[ \frac{\gamma - |h_{aw}|^2 P_\text{a} - |h_{bw}|^2 P_{\text{min}} -\sigma_w^2}{|h_{bw}|^2 \left(P_{\text{max}} - P_{\text{min}}  \right) } \right] ,
\end{aligned}
\end{equation}
which has an increasing partial derivative w.r.t $\gamma$, given by $\frac{\pi_1}{|h_{bw}|^2 \left(P_{\text{max}} - P_{\text{min}} \right)}$, thus $\gamma = |h_{bw}|^2 P_{\text{max}} + \sigma_w^2$ should be chosen.

\subsubsection*{(3) $|h_{bw}|^2 P_{\text{min}} + |h_{aw}|^2 P_\text{a}+\sigma_w^2 < \gamma \leq |h_{bw}|^2 P_{\text{max}}+\sigma_w^2$}
In this case, we have
\begin{equation}\label{3.16}
\begin{aligned}
\mathbb{P}_{E} &= \pi_0 \mathbb{P}_{FA} + \pi_1 \mathbb{P}_{MD} \\
&= \pi_0 \left[ \frac{|h_{bw}|^2 P_{\text{max}}+\sigma_w^2 -\gamma}{|h_{bw}|^2 \left(P_{\text{max}} - P_{\text{min}}  \right) } \right]  \\
& \hspace{1cm} +  \pi_1 \left[ \frac{\gamma - |h_{aw}|^2 P_\text{a} - |h_{bw}|^2 P_{\text{min}} -\sigma_w^2}{|h_{bw}|^2 \left(P_{\text{max}} - P_{\text{min}}  \right) } \right],
\end{aligned}
\end{equation}
and
\begin{equation}\label{3.17}
\begin{aligned}
\frac{\partial \mathbb{P}_{E}}{\partial \gamma} &= \frac{\pi_1-\pi_0}{|h_{bw}|^2 \left(P_{\text{max}} - P_{\text{min}}  \right)} =
\begin{cases}
\geq 0 , & \text{if} \quad  \pi_1 \geq \pi_0  \\
< 0 , & \text{otherwise}.
\end{cases}
\end{aligned}
\end{equation}
Based on the knowledge of $\pi_0$ and $\pi_1$, Willie can choose the optimal value of $\gamma$.

The corresponding $\mathbb{P}_E$ for the choice of optimal threshold, $\gamma^*$, can be found by using the appropriate expressions of $\mathbb{P}_E$ from Case-II, hence concluding the proof.

\end{document}